\documentclass[aps,prb,twocolumn,superscriptaddress,showpacs,10pt]{revtex4-1}
\usepackage{bm}					
\usepackage{graphicx}
\usepackage{amsmath,amsfonts,amssymb}	
\usepackage{latexsym} 				
\usepackage{xcolor}
\usepackage{hyperref}
\usepackage{tikz}

\begin{document}

\title{Signatures of Rashba Spin-Orbit Interaction in Charge and Spin Properties of Quantum Hall Systems}

\author{Daniel \surname{Hernang\'{o}mez-P\'{e}rez}}

\affiliation{Laboratoire de Physique et Mod\'elisation des Milieux Condens\'es, CNRS and Universit\'e Joseph Fourier, B.P. 166, 25 rue des Martyrs, F-38042 Grenoble, France}

\author{Serge Florens}
\affiliation{Institut N\'eel, CNRS and Universit\'e Joseph Fourier, B.P. 166, 25 rue des Martyrs, F-38042 Grenoble, France}

\author{Thierry Champel}

\affiliation{Laboratoire de Physique et Mod\'elisation des Milieux Condens\'es, CNRS and Universit\'e Joseph Fourier, B.P. 166, 25 rue des Martyrs, F-38042 Grenoble, France}

\date{\today}


\begin{abstract}
We study the local equilibrium properties of two-dimensional electron gases at high magnetic fields in the presence of random smooth electrostatic disorder, Rashba spin-orbit coupling, and the Zeeman interaction. Using a systematic magnetic length ($l_B$) expansion within a Green's function framework we derive quantum functionals for the local spin-resolved particle and current densities which can be useful for future studies combining disorder and mean-field electron-electron interaction in the quantum Hall regime. We point out that the spin polarization presents a peculiar spatial dependence which can be used to determine the strength of the Rashba coupling by local probes. The spatial structure of the current density, consisting of both compressible and incompressible contributions, also essentially reflects the effects of Rashba spin-orbit interaction on the energy spectrum. We show that in the semiclassical limit $l_B \rightarrow 0$ the local Hall conductivity remains, however, still quantized in units of $e^2/h$ for any finite strength of the spin-orbit interaction. In contrast, it becomes half-integer quantized when the latter is infinite, a situation which corresponds to a disordered topological insulator surface consisting of a single Dirac cone. Finally, we argue how to define at high magnetic fields a spin Hall conductivity related to a dissipationless angular momentum flow, which is characterized by a sequence of plateaus as a function of the inverse magnetic field (thus free of resonances).
\end{abstract}

\maketitle

\section{Introduction}\label{sec_introduction}
\subsection{Motivation}
Over the past few years, Rashba spin-orbit (SO) coupling\cite{Rashba1960, Rashba1984} has attracted a considerable amount of attention both from the experimental and theoretical perspectives with the objective to better understand  electrical and spin transport properties in disordered semiconductor systems at the nanoscale. \cite{Zutic2004} One of the main reasons for this keen interest is that, from a practical point of view, nanodevices built from these systems present a high potential for technological applications, since in low-dimensional heterostructures with asymmetrically confined two-dimensional electron gases (2DEGs) the interaction between the spin and the electric charge can be controlled by external local electric fields.\cite{Enoki1997} On the theoretical plane, the comprehension of these physical phenomena requires us to consider the complex dynamical interplay of charge and spin in confined nanostructures together with the motion of the carriers taking place in the presence of a disordered potential landscape. 

In this context, 2DEGs under high perpendicular magnetic fields may constitute one of the preferred playgrounds for the fundamental study of Rashba SO interaction since the effect of disorder can be well captured semiclassically\cite{Trugman1983, Prange1987, Fogler1997}, with an electronic charge motion resulting from the combination of a fast cyclotron motion and a slow guiding center drift in the presence of an arbitrary potential energy. At low temperatures, this characteristic motion is responsible for spectacular quantum signatures in macroscopic transport properties\cite{VonKlitzing1980} such as the well-known perfect quantization of the Hall conductance together with the simultaneous vanishing of the longitudinal conductance. The origin of the Hall plateaus can be found in the reservoir of localized states in the bulk of the system where the guiding center of the quantized cyclotron orbits drifts along quasiexclusively closed equipotential lines of the random  potential landscape, by reason of the ineffective energy exchange between the two (orbital and guiding) degrees of freedom. Simultaneously, macroscopic transport across the sample can only occur through the quasiballistic edge channels\cite{Buttiker1988}, as long as the chemical potential remains pinned away from the center of the disorder-broadened Landau levels which corresponds to the percolation threshold for the disorder equipotential lines. At the plateaus transitions, a diffusive electronic propagation throughout the system sets in via a bulk percolating network of equipotential lines, with the electron at a single energy passing through several saddle points of the potential\cite{Trugman1983} as recently evidenced experimentally by scanning tunneling spectroscopy measurements.\cite{Morgenstern2008}

Even though some  previous works\cite{Vasilopoulos2003, Shen2004, Shen2005,Reynoso2004, Kramer2005, Pala2005} have already considered transport properties of 2DEGs with strong magnetic fields in the presence of Rashba SO interaction, none of them specifically addresses the peculiar quantum Hall  transport regime. A few theoretical papers have studied the problem of charge transport either in bulk clean systems\cite{Vasilopoulos2003, Shen2004, Shen2005} or 2DEGs with simple confinement potential profiles\cite{Reynoso2004, Kramer2005, Pala2005} (quantum Hall edge states toy model) but, in general, little consideration has been given to the combined effects of Rashba SO coupling and Zeeman interaction (which also has to be included) on the electronic bulk drift states. One of the main reasons is that one has to deal at high magnetic fields with the nonperturbative quantum problem of lifting the huge Landau level degeneracy by long-range disorder, for which the usual techniques based on the formalism of perturbation theory turn out to be irrelevant. Another difficulty resides in the formulation of a consistent transport theory, which requires including Landau level mixing processes in the analysis, since the matrix elements of the current density necessarily relate adjacent Landau levels. In addition, macroscopic transport coefficients at high magnetic fields are dominated by long range spatial inhomogeneities. As a consequence, experimentally measured (macroscopic) conductances are no longer related to (microscopic) conductivities by simple dimensional factors given by the sample geometry\cite{Prange1987}, but present percolation features characterizing the fractal nature of the global transport network. In this line, it has also been suggested\cite{Lee1994,Hanna1995,Meir2002} that disordered quantum Hall systems with SO coupling may belong to a different universality class from spin-degenerate electron systems since they may present different critical exponent for the percolation correlation length under 
certain circumstances.

The understanding of the nontrivial physics of quantum Hall systems in the presence of Rashba and Zeeman couplings is also crucial to explore the spin (angular momentum) flow in this regime. It has been argued\cite{Shen2004, Shen2005, Bao2005} that the spin Hall conductance (appearing since a voltage drop can generate a transverse spin current in the presence of SO interaction) diverges in clean systems whenever two SO-split energy levels become accidentally degenerate. The same behavior has been predicted in other observables such as the spin polarization appearing as a response to an external electric field.\cite{Bao2007} Experimentally, this resonant behavior has still not been measured due to difficulties in discriminating the origin of the spin accumulation\cite{Kato2004} (thus the type of SO interaction) and to the lack of experimental probes that couple to the spin current. To make matters worse from a theoretical perspective, the spin Hall current used in the former works\cite{Shen2004, Shen2005, Bao2005} does not satisfy a continuity equation\cite{Niu2006} (which clearly makes the concept of spin accumulation at the sample edges problematic) and we have to include the so-called spin torque contribution as a source term to preserve the spin density. This is a normal consequence of the presence of SO interaction, since the spin is not a good quantum number. However, the spin current defined in this way (and therefore the spin Hall conductance) is not uniquely determined since we can add to it any current, at the price of a modification in the spin torque term, still verifying the conservation laws. To avoid any ambiguity, it is therefore conceptually more natural to define a SO-polarized current along the spin precession axis, an approach that will be presented in this work.

\subsection{Organization of the paper}

In this paper we develop an analytical theory of electronic transport for disordered quantum Hall systems in the hydrodynamic (local equilibrium) regime taking into account both Rashba and Zeeman interactions. Since this work uses original techniques based on phase space quantization\cite{Zachos2005}, we give beforehand a brief description of our main results. First, in Sec. \ref{sec_equation_of_motion}, we present the SO vortex Green's function formalism. We show that the single-particle Green's function can be found using a systematic expansion in powers of the magnetic length. In Sec. \ref{sec_electron_density}, we derive a general quantum-mechanical expression for the local spin-resolved electron density valid at high magnetic fields, which takes the form of a functional of an (arbitrary) potential energy.
We provide in Sec. \ref{sec_electron_current} a detailed derivation of the local electron current density, discuss both the density-gradient (edge) and drift (bulk) contributions, and explain their spatial dispersion in the presence of disorder and / or confinement potentials (i.e., quantum wires under high magnetic fields). Furthermore, in Sec. \ref{subsec_nonequilibrium} we consider the nonequilibrium current density in the linear response regime to compute an explicit formula for the local conductivity tensor in the presence of Rashba SO and Zeeman interactions. We prove that the Hall (off-diagonal) component of the tensor presents preserved quantized plateaus with average value given by integer units of the conductance quantum, $e^2/h$, that are robust to finite Rashba coupling. Our general formula is also used to obtain the local Hall conductivity in a disordered topological insulator film (in the limit of vanishing coupling between top and bottom surfaces) at high fields, explicitly proving from a microscopic derivation its half-integer quantization. We also point out how to define a local spin Hall conductivity that has a physical meaning in the presence of SO interaction. This observable is shown to give a dissipationless angular momentum flow presenting fingerprints of the particle spin imbalance below the chemical potential. We shall finally  present in this section the derivation of the macroscopic transport coefficients in a high-temperature regime, and  demonstrate that the longitudinal conductance is characterized by a classical percolation exponent, which remains unaffected by the presence of Rashba SO coupling. A general conclusion describing future directions of this work closes the paper, together with some calculation details given in the Appendix.

\section{Equation of motion in the spin-orbit vortex representation}\label{sec_equation_of_motion}
\subsection{Spin-orbit vortex states}\label{sec_spin_vortex}
We now proceed to the theoretical analysis for arbitrary smooth disorder potentials, which allows great simplifications at high magnetic fields. The SO vortex states, on which our quantum high magnetic field theory relies, are eigenstates of the clean Hamiltonian 
\begin{equation}\label{Hamiltonian_clean}
\hat{\mathcal{H}}_0 = \hat{\mathcal{H}}_{\textnormal{2DEG}} + \hat{\mathcal{H}}_{\textnormal{R}} + \hat{\mathcal{H}}_\textnormal{Z},
\end{equation}
which describes a single electron of charge $e = -|e|$, effective mass $m^\ast$, and spin $s = 1/2$ confined in a two-dimensional (2D) plane, under a uniform perpendicular magnetic field $\mathbf{B} = B\hat{\mathbf{z}}$ and in the presence of both Rashba SO and Zeeman interactions. The first term in Eq. \eqref{Hamiltonian_clean} is the spin-diagonal Hamiltonian of a free single electron
\begin{equation}\label{Hamiltonian_2DEG}
 \hat{\mathcal{H}}_{\textnormal{2DEG}} = \dfrac{\hat{\bm{\Pi}}^2}{2m^\ast} \otimes 1\!\!1_s,
\end{equation}
with 
\begin{equation}
 \bm{\Pi} =\mathbf{p} - \dfrac{e}{c} \mathbf{A}(\mathbf{r})
\end{equation}
the gauge-invariant (kinetic) momentum [here $\mathbf{p}$ is the canonical momentum, $\mathbf{A}(\mathbf{r})$ the electromagnetic vector potential with $\mathbf{r} = (x,y)$ the position of the electron in the 2D plane and $c$ the speed of light], $1\!\!1_s$ the identity matrix in spin space, and $\otimes$ the tensor product symbol. The next term is the so-called Rashba Hamiltonian\cite{Rashba1960, Rashba1984} describing the coupling between orbital and spin motions 
\begin{equation}\label{Hamiltonian_Rashba}
 \hat{\mathcal{H}}_{\textnormal{R}} = \alpha [\hat{\Pi}_x \otimes \sigma_y - \hat{\Pi}_y \otimes \sigma_x],
\end{equation}
with $\alpha$ a spatially constant Rashba SO parameter (for simplicity, we do not consider here random spatial fluctuations of the Rashba coupling field) and $\sigma_\gamma$ ($\gamma\in \{x,y\}$) the Pauli matrices. Finally, the last term represents the Zeeman interaction between the electron's intrinsic magnetic moment and the external magnetic field
\begin{equation}\label{Hamiltonian_Zeeman}
  \hat{\mathcal{H}}_{\textnormal{Z}}=\dfrac{1}{2}g \mu_{B} B \otimes\sigma_{z},
\end{equation}
where $g$ is the Land\'e $g$ factor\cite{footnote_g} and $\mu_B = |e|\hbar/(2m_0c)$ the Bohr's magneton with $m_0$ the bare electron mass.

The energy spectrum corresponding to the Hamiltonian \eqref{Hamiltonian_clean} is composed by discrete SO-split Rashba-Landau levels\cite{Rashba1984} and reads
\begin{equation}\label{spectrum}
E_{n,\lambda} = \hbar \omega_c \left(n -\dfrac{\lambda}{2} \Delta_n \right),
\end{equation}
where $\omega_c = |e|B/(m^\ast c)$ is the cyclotron pulsation and $\Delta_n =\sqrt{(1 - Z)^2 + nS^2}$ is a dimensionless quantity defining the gap between the nonequidistant SO-split energy levels in units of the cyclotron energy $\hbar \omega_c$. The multiple level crossings appearing in the spectrum are controlled by the parameters $S$ and $Z$, measuring the relative strength of the Rashba SO coupling [per magnetic length, $l_B = \sqrt{\hbar c / (|e|B)}$] and Zeeman interaction\cite{footnote1} to the cyclotron energy. They are defined by the following expressions:
\begin{equation}
 S =\dfrac{\alpha 2 \sqrt{2}}{\omega_c l_B}, \label{S} \hspace*{1.cm}
 Z = \dfrac{g}{2} \dfrac{m^{\ast}}{m_0}.
\end{equation}
It is clear that the energy levels in Eq. \eqref{spectrum} are highly (macroscopically) degenerate since they depend on just two quantum numbers (while three are required to fully represent the quantum states): a positive integer $n \geq 0$, called the Rashba-Landau level index, and $\lambda = \lambda(n)$, the SO quantum number which is a function of the former and takes the values $\lambda = \pm$ if $n \neq 0$ and $\lambda = -$ otherwise. An adequate treatment of the lifting of the Rashba-Landau level degeneracy by a random potential energy, within the characteristic quantum Hall regime at high magnetic fields, is therefore clearly needed.

Working in the symmetrical gauge $\mathbf{A}(\mathbf{r}) = \mathbf{B} \times \mathbf{r}/2$, a very convenient basis of eigenfunctions of the Hamiltonian operator \eqref{Hamiltonian_clean} is the set of semiorthogonal SO vortex wave functions\cite{Hernangomez2013} with real-space representation $\tilde{\Psi}_{n,\lambda,\mathbf{R}}(\mathbf{r}) \equiv \langle \mathbf{r}| n, \lambda, \mathbf{R} \rangle$,
\begin{equation}\label{SOvortex_states}
 \tilde{\Psi}_{n,\lambda,\mathbf{R}}(\mathbf{r}) = \sum_{\sigma=\pm}f_\sigma(\theta_n^{\lambda})\Psi_{n_\sigma,\mathbf{R}}(\mathbf{r}) \otimes |\sigma \rangle.
\end{equation}
These eigenfunctions are labeled by the collection of quantum numbers $\nu = \{n,\lambda, \mathbf{R}\}$ where $\mathbf{R} =(X,Y)$ is a doubly continuous quantum number corresponding to the position in the plane of the center of the localized (vortex-like) spinor wave function. The spinor corresponds to a very particular combination of the eigenstates of the Pauli matrix $\sigma_z$, i.e., $\sigma_z |\sigma \rangle = \sigma |\sigma \rangle$, with a spatial part characterized by the so-called vortex states\cite{Malkin1969, Champel2007},
\begin{multline}\label{vortex_states}
  \Psi_{n,\mathbf{R}}(\mathbf{r})=\dfrac{1}{\sqrt{2 \pi l_{B}^{2} n!}} 
\left[\dfrac{x-X+i(y-Y)}{\sqrt{2}l_{B}} \right]^{n} \\ \times  
\exp \left[-\dfrac{(x-X)^2+(y-Y)^2+2i(yX-xY)}{4l_{B}^{2}}\right].
\end{multline}
We note that each of the vortex wave functions appearing in the components $\sigma=\pm$ of the spinor \eqref{SOvortex_states} is labeled by a spin-dependent Landau level quantum number defined by $n_\sigma = n - (1+\sigma)/2$ and weighted by the function
\begin{equation}
 f_\sigma(\theta_n^{\lambda})=
 \begin{cases}
 \sin (\theta_n^\lambda) &\sigma=+,\\
\cos (\theta_n^\lambda) & \sigma=-.
\end{cases}
\end{equation}
As such, the wave functions \eqref{SOvortex_states} are characterized by a probability density composed by two unequal maxima localized along adjacent spin-resolved Landau orbits whose different cyclotron radii collapse into a single one when the SO parameter $S$ vanishes. Here, we have set $\Psi_{-1,{\bf R}}({\bf r}) \equiv 0$ so that the above formula \eqref{SOvortex_states} also holds for the lowest Landau level $n=0$.

The angular parameters $\theta_n^{\lambda}$ are defined by\cite{footnote2}
\begin{equation}\label{angles}
 \theta^{\lambda}_{n}= \arctan \left[ \dfrac{(1-Z)+\lambda \Delta_n}{S \sqrt n} \right].
\end{equation}
For $n\geq 1$, both projections along the Rashba dependent spin axis satisfy the identity 
\begin{equation}
 \theta_n^+ = \theta_n^{-} + \pi/2,
\end{equation}
ensuring that SO vortex states with the same Rashba-Landau level index $n$ but opposite SO quantum number $\lambda$ are orthogonal. Note that two SO vortex states with the same $n$ and $\lambda$ but different guiding center positions ${\bf R}$ present a nonzero overlap characteristic of the coherent states algebra. Despite this nonorthogonality, the full set of states $\{|\nu \rangle \} = \{|n,\lambda, \mathbf{R}\rangle\}$ does form a basis, as is well known for coherent states; see e.g.,  Ref. \onlinecite{Champel2007} and references therein. 

\subsection{Dyson equation}
We have previously shown\cite{Hernangomez2013} that the full electronic Green's function, corresponding to the Hamiltonian operator
\begin{equation}\label{Hamiltonian}
 \hat{\mathcal{H}} = \hat{\mathcal{H}}_0 + V(\hat{\mathbf{r}}) \otimes 1\!\!1_s,
\end{equation}
where $\hat{\mathcal{H}}_0$ is the clean Hamiltonian given in Eq. \eqref{Hamiltonian_clean} and $V(\hat{\mathbf{r}}) \otimes 1\!\!1_s$ is an electrostatic potential energy term diagonal in spin space, can be written exactly using the SO vortex wave functions as 
\begin{multline}\label{electronic_Green_function} 
 G^{R,A}_{\sigma\sigma'}(\mathbf{r},\mathbf{r}',\omega)=\int\dfrac{d^2\mathbf{R}}{2 \pi l_B^2} \sum_{n_1,\lambda_{1}}\sum_{n_2,\lambda_{2}} f_{\sigma}(\theta_{n_1}^{\lambda_{1}}) f_{\sigma'}(\theta_{n_2}^{\lambda_{2}}) \\  \times  K_{n_{1\sigma};n_{2\sigma'}}(\mathbf{r},\mathbf{r}',\mathbf{R})\,{g}^{R,A}_{n_{1},\lambda_{1};n_{2},\lambda_{2}}(\mathbf{R},\omega).  
\end{multline}
Here $R, A$ respectively stand for the retarded and advanced components of the Green's function expressed in the position representation, $G(\mathbf{r},\mathbf{r}',\omega) = \langle \mathbf{r}|\hat{G}(\omega)|\mathbf{r}' \rangle$, and $\sigma,\sigma' \in \{\pm\}$ define its matrix elements in spin space. The electronic structure factor $K_{n_1;n_2}(\mathbf{r},\mathbf{r}' ,\mathbf{R})$, which contains all the information about the electronic wave functions, is defined by 
\begin{eqnarray}\label{kernel_function} 
K_{n_1;n_2}(\mathbf{r},\mathbf{r}',\mathbf{R})= {\rm e}^{(-l_B^2/4)\Delta_\mathbf{R}}\left[\Psi_{n_{2},\mathbf{R}}^\ast(\mathbf{r}')\Psi_{n_{1},\mathbf{R}}(\mathbf{r})\right], \hspace{0.5cm}
\end{eqnarray} 
with $\Delta_\mathbf{R}$ the Laplacian operator taken with respect to the 2D vortex position ${\bf R}$. This function plays the role of a convolution kernel, independent of the (arbitrary) potential energy $V(\mathbf{r})$, and gives the quantum contribution to the orbital motion of the electron in the $2$D plane. In this sense, Eq. \eqref{electronic_Green_function} can be understood as the quantum formulation of the decomposition of the electronic motion into orbital and guiding center (or vortex) degrees of freedom, the latter being characterized by the functions $g^{R,A}_{n_1, \lambda_{1};n_2,\lambda_{2}}(\mathbf{R},\omega)$ which have still to be determined.

The local SO vortex Green's function $g^{R,A}_{n_1, \lambda_{1};n_2,\lambda_{2}}(\mathbf{R},\omega)$ can be found to obey the following exact equation of motion\cite{Hernangomez2013}
\begin{multline}\label{Dyson_star_product} 
  (\omega-E_{n_1,\lambda_1} \pm i0^+) {g}^{R,A}_{n_1,\lambda_{1};n_2,\lambda_{2}} (\mathbf{R},\omega) = \delta_{n_1,n_2}\delta_{\lambda_{1},\lambda_{2}} \\ + \sum_{n_3,\lambda_{3}} v_{n_1,\lambda_{1};n_3,\lambda_{3}}(\mathbf{R}) \star g^{R,A}_{n_3,\lambda_{3};n_2,\lambda_{2}}(\mathbf{R},\omega), 
\end{multline}
where the retarded (resp. advanced) SO vortex Green's function corresponds to the plus (resp. minus) sign and $0^{+}$ is a positive infinitesimal quantity encoding the information related to causality. The symbol $\star$ is a magnetic version of the Groenewold-Moyal $\star$ product\cite{Zachos2005}, a pseudo-differential infinite-order symplectic operator given by 
\begin{equation}\label{starproduct} 
\star = \exp{\left[i \dfrac{l^{2}_{B}}{2}(\overleftarrow{\partial}_{X}  \overrightarrow{\partial}_{Y}-\overleftarrow{\partial}_{Y} \overrightarrow{\partial}_{X}) \right]}.
\end{equation}
The arrow above each of the partial derivatives indicates to which side (left/right) it has to be applied. The non-commutative $\star$ product condenses the dynamics in the phase space defined by the conjugated vortex variables $(X,Y)$, which for 2D electrons under a perpendicular magnetic field corresponds to the physical space for the components of the guiding center position $\mathbf{R}$. In this case, the squared magnetic length $l_B^2$ plays the role of a deformation parameter that allows one to smoothly recover (semi)classical dominant terms for the guiding center dynamics from the continuous limit $l_B \rightarrow 0$.

In the local equilibrium regime, the potential operator $V(\hat{\mathbf{r}}) \otimes 1\!\!1_s$  will contain two different types of terms: first, an effective electrostatic contribution $V_{\textnormal{eff}}(\mathbf{r})$ resulting from  random impurity potentials, confinement and/or electron-electron repulsion taken into account at the mean-field level; and second, an external contribution proportional to a spatially varying electrochemical potential $\Phi(\mathbf{r})$ applied to the sample:
\begin{equation}\label{potential_energy}
V(\mathbf{r}) = V_{\textnormal{eff}}(\mathbf{r}) + e \Phi(\mathbf{r}).
\end{equation}
The electrochemical potential (containing both contributions from the macroscopic electric field and the macroscopic chemical potential gradients) can be related to a macroscopic electromotive field $\mathbf{E}(\mathbf{r})= -\nabla_{\mathbf{r}}\Phi(\mathbf{r})$ that induces in the system a nonequilibrium response as a consequence of a macroscopic voltage drop.

The matrix elements $v_{n_1,\lambda_{1};n_2,\lambda_{2}}(\mathbf{R})$ appearing in Dyson equation \eqref{Dyson_star_product} play the role of an effective potential energy seen by the vortex and  can be evaluated exactly in the SO vortex representation\cite{Hernangomez2013} 
\begin{equation}\label{potential_modified}
 v_{n_1,\lambda_{1};n_2,\lambda_{2}}(\mathbf{R}) = \sum_{\sigma=\pm}f_\sigma(\theta_{n_1}^{\lambda_1}) f_{\sigma}(\theta_{n_2}^{\lambda_2})v_{n_{1\sigma};n_{2\sigma}}(\mathbf{R}),
 \end{equation}
 with the reduced matrix elements  $v_{n_{1\sigma};n_{2\sigma}}(\mathbf{R})$ dependent on the spin projection given by
 \begin{equation}\label{potential_modified_spinless}
v_{n_{1\sigma}; n_{2\sigma}}(\mathbf{R}) = \int d^2 \bm{\eta}\,K_{n_{1\sigma};n_{2\sigma}}(\bm{\eta},\bm{\eta},\mathbf{0}) V(\bm{\eta} + \mathbf{R}). 
 \end{equation}
These reduced matrix elements, expressed as a convolution of the bare potential energy $V({\bf r})$ with the electronic kernel \eqref{kernel_function} encoding the information about the cyclotron motion, can obviously be physically interpreted as an average of the bare electrostatic potential $V({\bf r})$ over the (integrated out) fast orbital angular degree of freedom.

\subsection{Spin-orbit vortex Green's functions at high magnetic fields}\label{subsec_SO_Green}
Finding an analytical exact solution of Dyson equation for an arbitrary $2$D potential $V(\mathbf{r})$ and all possible values of the external magnetic field is unquestionably an overwhelming task. However, in this work  we are only interested in the high magnetic field regime where the magnetic length $l_B$ becomes the smallest length scale (with respect to the characteristic length scale related to the local variations of the smooth potential energy).
This suggests that one possible strategy to follow\cite{Champel2008} in order to obtain a solution to the equation of motion in the presence of arbitrary smooth disorder potentials and macroscopic nonequilibrium electromotive fields is to expand the matrix elements of the potential given in Eq. \eqref{potential_modified_spinless} in a power series in (the small parameter) $l_B$,
\begin{equation}\label{potential_expansion_full}
{v}_{n_{1\sigma}; n_{2\sigma}}(\mathbf{R}) = \sum_{j=0}^{+\infty}\left(\dfrac{l_B}{\sqrt{2}}\right)^j {v}^{(j)}_{n_{1\sigma}; n_{2\sigma}}(\mathbf{R}),
\end{equation}
with ${v}^{(j)}_{n_{1\sigma}; n_{2\sigma}}(\mathbf{R})$ independent of $l_B$, and to find the SO vortex Green's functions by writing them analogously, 
\begin{equation}\label{Green_SO_expansion}
 g_{n_1,\lambda_1;n_2,\lambda_2}(\mathbf{R}, \omega) = \sum_{j=0}^{+\infty} \left(\dfrac{l_B}{\sqrt{2}}\right)^j g^{(j)}_{n_1,\lambda_1;n_2,\lambda_2}(\mathbf{R},\omega).
\end{equation}
Here, and subsequently, we drop the retarded and advanced superscripts, since the corresponding SO vortex Green's functions can be easily identified from the sign of their imaginary parts. The functions $g^{(j)}_{n_1,\lambda_{1};n_2,\lambda_{2}}(\mathbf{R},\omega)$ appearing in the series expansion of the SO vortex Green's function can then be computed solving iteratively the equation of motion \eqref{Dyson_star_product} order by order in powers of the magnetic length. This procedure allows us in turn to systematically obtain  quantum microscopic expressions (at finite $l_B$) for any local observable in the hydrodynamic regime, under the form of functionals of the arbitrary potential energy $V({\bf r})$.

Before providing the resulting expressions for the SO vortex Green functions, a few comments are in order. It is important to note that the Green's function decomposition \eqref{electronic_Green_function} of the electronic motion into a slow guiding center drift and a fast cyclotron motion reveals that the full dependence on the magnetic length of the electronic Green's function has two distinct origins related to each of the coupled motions. The $l_B$ expansion discussed here only affects the vortex contribution to the electronic Green's function, so that the computed formulas preserve through the structure factors some quantum-mechanical features such as wave function spreading and interference effects associated with the quantized cyclotron motion. Further expansion in $l_B$ of these electronic kernels allows one to recover entirely semiclassical expressions for the observables, where only the quantization of the kinetic energy is kept intact while the full internal spatial structure resulting from quantum interferences is neglected. 

This distinction in the $l_B$ contributions  is, however, rather subtle, since one has in principle the freedom to modify the $\star$ product with a trivial rotation in the vortex phase space. For instance, one can  make integrations by parts in Eq. \eqref{electronic_Green_function}, so that the differential operator ${\rm e}^{-(l_B^2/4) \Delta_{{\bf R}}}$ appearing in expression \eqref{kernel_function} is then included into a new guiding center contribution, which will obey a Dyson equation analogous to Eq. \eqref{Dyson_star_product} with a new infinite-order differential operator product. In this case, the structure factor for the orbital motion is simply given by the product $\Psi_{n_{2},\mathbf{R}}^\ast(\mathbf{r}')\Psi_{n_{1},\mathbf{R}}(\mathbf{r})$ of two (localized) vortex wave functions. Clearly, this phase space rotation corresponds to an infinite-order resummation of a whole class of $l_B$-dependent terms. It has been proved in previous works\cite{Champel2009,Champel2010} that the structure factor \eqref{kernel_function} for the orbital motion is preferentially selected as early as the Landau level degeneracy is lifted, essentially due to the specific structure of the $\star$ product operator. Quite remarkably, the $\star$ product indeed generates\cite{Champel2009,Champel2010} within the guiding center dynamics a hierarchy of local characteristic energy scales built from the spatial derivatives of the effective potentials and corresponding to intrinsic invariants, such as the local Gaussian curvature. For an arbitrary 2D potential $V(\mathbf{R})$ this means that at leading order one can safely replace the $\star$ product with the usual product between functions, as long as the thermal energy scale exceeds the curvature energy scale. This leads to a considerable simplification since the system of partial differential equations obeyed by the SO vortex Green's functions then transforms into a system of algebraic equations. From the physical point of view, the existence of a hierarchy of energy scales expresses a stability property (or robustness) of the chosen representation of quantum states.

Because the $\star$ product  involves products of derivatives of the vortex coordinates acting along orthogonal directions, it is obvious that this formalism becomes exact for any 1D potential. This pinpoints the quantum-mechanical effects encapsulated within the structure factor, namely, the dragging of the nodal pattern related to the quantized orbital motion along the guiding center trajectory. In other terms, the operator  ${\rm e}^{-(l_B^2/4) \Delta_{{\bf R}}}$ in Eq. \eqref{kernel_function} realizes the 1D delocalization of the vortex states, which are originally localized in any direction. Our quantum theory somehow corresponds to a generalization of the quantum drift state picture pioneered by Trugman\cite{Trugman1983} thirty years ago. 

We now come back to the derivation of the SO vortex Green's functions. The equation of motion satisfied by the leading-order component of the SO vortex Green's function  can be trivially found setting $j=0$ in Eqs. \eqref{potential_expansion_full} and \eqref{Green_SO_expansion}. At this order, we can easily see that the matrix elements of the potential are diagonal both in the Rashba-Landau level index and the SO quantum number
\begin{equation}
 v^{(0)}_{n_1,\lambda_1;n_2,\lambda_2}(\mathbf{R}) = \delta_{n_1,n_2}\delta_{\lambda_1,\lambda_2}V(\mathbf{R}),
\end{equation}
and Dyson equation \eqref{Dyson_star_product} takes a closed form. The SO vortex Green's functions present then a well-known simple pole structure
\begin{equation}\label{Green_SO_zero}
 g^{(0)}_{n_1,\lambda_{1};n_2,\lambda_{2}}(\mathbf{R},\omega) = \dfrac{\delta_{n_1,n_2} \delta_{\lambda_{1},\lambda_{2}}}{\omega - \xi_{n_1,\lambda_{1}}(\mathbf{R}) \pm i0^+}, 
\end{equation}
where the energies read $\xi_{n,\lambda}(\mathbf{R})=E_{n,\lambda} + V(\mathbf{R})$. 
We see that the main effect of the potential energy  is to locally lift the macroscopic degeneracy of the Rashba-Landau energy levels with respect to the guiding center coordinates $\mathbf{R}$, while keeping $n$ and $\lambda$ as good quantum numbers. 

It is anticipated that the leading-order SO vortex Green's functions  do not contribute to drift transport, since the latter is related to the gradient of the potential energy. In other terms, the drift motion  necessarily involves, up to some extent, Rashba-Landau level mixing processes which are subdominant at high magnetic fields, so that we have to carry on our analysis by including\cite{Champel2007} the first $l_B$ corrections to the SO vortex Green's functions. The subleading term [linear in $l_B$ within expansion \eqref{Green_SO_expansion}] is obtained from the algebraic equation
\begin{multline}
 \left[\omega - \xi_{n_1,\lambda_{1}}(\mathbf{R}) \pm i0^{+}\right] g^{(1)}_{n_1,\lambda_{1};n_2,\lambda_{2}}(\mathbf{R},\omega) =\\ \sum_{n_3,\lambda_{3}} v^{(1)}_{n_1,\lambda_{1};n_3,\lambda_{3}}(\mathbf{R}) g^{(0)}_{n_3,\lambda_{3};n_2,\lambda_{2}}(\mathbf{R},\omega),
\end{multline}
whose solution reads
\begin{multline}\label{Green_SO_first}
 g^{(1)}_{n_1,\lambda_1;n_2,\lambda_2}(\mathbf{R},\omega) =  \\
  \dfrac{v^{(1)}_{n_1,\lambda_1;n_2,\lambda_2}(\mathbf{R})}{[\omega -\xi_{n_1,\lambda_1}(\mathbf{R}) \pm i0^+][\omega -\xi_{n_2,\lambda_2}(\mathbf{R}) \pm i0^+]}.
\end{multline}
The first-order contribution to the SO vortex Green's function therefore presents a familiar structure with two poles. The subdominant potential matrix elements are given by
\begin{eqnarray}\label{potential_first_order}
 v^{(1)}_{n_1,\lambda_{1};n_2,\lambda_{2}}(\mathbf{R}) &=& \sum_{\sigma=\pm} f_\sigma(\theta_{n_1}^{\lambda_{1}})f_\sigma(\theta_{n_2}^{\lambda_{2}})  \Big[\sqrt{n_{1\sigma} + 1}\notag \\ &\times&\delta_{n_{1\sigma} +1,n_{2\sigma}}\partial_{+}V(\mathbf{R}) + \textnormal{c.c.}\,(1\leftrightarrow 2)\Big],\hspace{0.5cm}
\end{eqnarray}
where $\partial_\pm = \partial_X \pm i \partial_Y$ and the notation $\textnormal{c.c.}\,(1\leftrightarrow 2)$ means exchanging indices and taking the complex conjugation in the former expression. As a difference to the dominant contribution, these subleading terms in the potential expansion induce a mixing between adjacent Rashba-Landau levels $n$ and different values of the SO quantum number $\lambda$ in the presence of a nonzero gradient of the smooth arbitrary potential. We would like to point out that the former recursive procedure can be done in principle up to any order in the $l_B$ expansion. However, higher order contributions in $l_B$ only lead to weak quantitative corrections for potentials which are smooth at the scale of the cyclotron radius, so they will be disregarded in the rest of this paper.

\section{Electron Density and Spin Polarization}\label{sec_electron_density}
\subsection{Quantum expression}
As a first simple step, we shall consider the local electron density. The local spectral density at equilibrium can be computed from the lesser component of the electronic Green's function evaluated at coinciding electron positions $\mathbf{r} = \mathbf{r}'$,
\begin{equation}\label{density_Glesser}
 n(\mathbf{r},\omega) = \textnormal{Tr}\left\{-iG^{<}(\mathbf{r},\mathbf{r},\omega) \right\},
\end{equation}
and can be written as a sum of two spin-resolved components (in the $\hat{\mathbf{z}}$ direction perpendicular to the $2$D plane)
\begin{equation}\label{density_full}
 n(\mathbf{r},\omega)= \sum_{\sigma=\pm} n_{\sigma}(\mathbf{r},\omega),
\end{equation}
with $n_\sigma(\mathbf{r},\omega)=-i G_{\sigma\sigma'}^{<}(\mathbf{r},\mathbf{r},\omega)\delta_{\sigma,\sigma'}$. The lesser component of the Green's function is obtained from the standard equilibrium relation 
\begin{equation}\label{Glesserdefinition}
 -iG^{<}(\mathbf{r},\mathbf{r},\omega)=in_{\textnormal{F}}(\omega)[G^{R}(\mathbf{r},\mathbf{r},\omega)-G^{A}(\mathbf{r},\mathbf{r},\omega)],
\end{equation}
where 
\begin{equation}\label{FermiDirac}
 n_{\textnormal{F}}(\omega)=\dfrac{1}{1+\exp[(\omega-\mu)/k_B T]}
\end{equation}
is the Fermi-Dirac distribution function at temperature $T$ and constant electrochemical potential (for global equilibrium) $\mu = e\Phi$ (we also designate by $k_B$ the Boltzmann constant). Inserting Eq. \eqref{electronic_Green_function} into \eqref{Glesserdefinition} and  using afterwards Eq. \eqref{density_Glesser}, we get the general formula for the local electronic density per spin after an integration over the energies:
\begin{align}\label{density_per_spin_general}
 n_{\sigma}(\mathbf{r}) &= -i \int\dfrac{d\omega}{2\pi}\int \dfrac{d^2 \mathbf{R}}{2 \pi l_B^2} \sum_{n_1,\lambda_{1}}\sum_{n_2,\lambda_{2}} f_{\sigma}(\theta_{n_1}^{\lambda_{1}}) f_{\sigma}(\theta_{n_2}^{\lambda_{2}})\notag \\ &\times K_{n_{1\sigma};n_{2\sigma}}(\mathbf{r},\mathbf{r},\mathbf{R})\, g^{<}_{n_1,\lambda_{1};n_2,\lambda_{2}}(\mathbf{R},\omega).
\end{align}
Here, the SO vortex lesser contribution $g^{<}(\mathbf{R},\omega)$ is defined in terms of the retarded and advanced components in the same way as the full electronic lesser Green's function $G^{<}(\mathbf{r},\mathbf{r},\omega)$. Equation \eqref{density_per_spin_general} also allows us to easily obtain the spin polarization in the direction perpendicular to the 2DEG, since the latter is proportional to the difference between spin-up and spin-down electronic populations:
\begin{equation}\label{polarization} 
 \Pi^z(\mathbf{r})=\dfrac{\hbar}{2}\left[n_{+}(\mathbf{r})-n_{-}(\mathbf{r})\right].
\end{equation}

At leading order in the series expansion in powers of the magnetic length the lesser component of the SO vortex Green's function, resulting from Eqs. \eqref{Green_SO_zero} and \eqref{Glesserdefinition}, is given by
\begin{multline}\label{Green_zero_lesser}
g_{n_1,\lambda_{1};n_2,\lambda_{2}}^{<\,(0)}(\mathbf{R},\omega) = 2 \pi in_{\textnormal{F}}(\omega)\delta_{n_1,n_2}\delta_{\lambda_{1},\lambda_{2}} \\ \times \delta[\omega-\xi_{n_1,\lambda_{1}}(\mathbf{R})].
\end{multline}
The dominant contribution to the spin-polarized electron density is then obtained after performing the straightforward summation over the energies in Eq. \eqref{density_per_spin_general} and reads
\begin{multline}\label{density_per_spin_leading}
 n_\sigma^{(0)}(\mathbf{r}) = \dfrac{1}{2} \int \dfrac{d^2 \mathbf{R}}{2 \pi l_B^2} \sum_{n=0}^{+\infty} \sum_{\lambda}  n_{\textnormal{F}}[\xi_{n,\lambda}(\mathbf{R})] \\ \times K_{n_\sigma}(\mathbf{r}-\mathbf{R}) \left[ 1 + \sigma \lambda \sqrt{1 - \dfrac{nS^2}{\Delta^2_n}} \right],
\end{multline}
with the diagonal elements of the electronic kernels $K_n(\mathbf{r}-\mathbf{R}) = K_{n;n}(\mathbf{r},\mathbf{r},\mathbf{R})$ given by\cite{Champel2009}
\begin{align}
K_{n}(\mathbf{r}-\mathbf{R})&= {\rm e}^{-(l_B^2/4)\Delta_\mathbf{R}}|\Psi_{n,\mathbf{R}}(\mathbf{r})|^2, \\
&=\dfrac{(-1)^n}{\pi l_B^2 } 
L_{n}\left[ \dfrac{2 ({\bf r}-{\bf R})^2}{l_B^2}\right] 
\, {\rm e}^{  -({\bf r}-{\bf R})^2/ l_B^2}.
\end{align}
We remind the reader that, for compatibility with the definition of the SO vortex wave functions, we take by convention $K_{-1}({\bf r}-{\bf R}) \equiv 0$.

We would like to stress again that Eq. \eqref{density_per_spin_leading} is far more general than  any semiclassical expression that can be derived in the strict limit $l_B\rightarrow 0$, since it still accounts for wave function broadening and electron delocalization encapsulated in the electronic kernels convoluted with the thermal vortex spectral density. More precisely, this quantum expression  \eqref{density_per_spin_leading} is expected to be accurate in a very wide regime of temperatures, as long as the thermal energy scale $k_B T$ overtakes the (quite small) local curvature characteristic energy scale. \cite{Champel2009,Champel2010}
From this expression it is not difficult to be convinced that the two spin-projected densities will contribute to the spatial dependence of the full electron density \eqref{density_full} with unequal weights, as a result of the SO Rashba interaction. This effect, notifying us that the spin is no longer a good quantum number, can be traced back to the particular entangled structure of the SO vortex wave functions \eqref{SOvortex_states} whose two spatial components mix adjacent energy level indices $n$, each one corresponding to different spin-polarized Landau levels.

\subsection{Semiclassical expression}

At high enough temperatures such that $k_B T \gg l_B |\bm{\nabla}_\mathbf{r} V(\mathbf{r})|$, we can make the replacement $\mathbf{R} \simeq \mathbf{r}$ inside the (smooth) functionals of the potential energy and integrate the vortex dependence taking into account the normalization condition of the electronic kernels $\int d^2 \mathbf{R} \,K_{n}(\mathbf{R})=1$. This is equivalent to consider a semiclassical regime, where the diagonal elements of the electronic kernels are replaced by 2D Dirac delta functions $K_n(\mathbf{r}-\mathbf{R})  \simeq \delta^{(2)}(\mathbf{r}-\mathbf{R})$. The quantum non-local expressions involving the vortex guiding center then transform into purely local functionals.
We obtain the following simple semiclassical expression for the particle density
\begin{equation}\label{density_full_scl}
 n_{\textnormal{sc}}(\mathbf{r}) = \dfrac{1}{2 \pi l_B^2} \sum_{n=0}^{+\infty} \sum_{\lambda} n_{\textnormal{F}}[\xi_{n,\lambda}(\mathbf{r})].
\end{equation}
Written in this form, it is also clear that we can identify $\lambda$ with a dressed spin  quantum number: both $\lambda$ projections contribute to the density with equal unitary weights as it happens with the real spin $\sigma$ for vanishing SO coupling. Note that this property also holds in the quantum realm [on the basis of Eq. \eqref{density_per_spin_leading}] as long as  $\lambda$ can be considered a good quantum number. In this sense, we can understand the high magnetic field result for the local electron density as being robust with respect to Rashba SO interaction, the only trace of SO coupling appearing in the particular energy spectrum \eqref{spectrum} inside the argument of the Fermi-Dirac distribution function that controls the width of the density plateaux.

\begin{figure}[tb] 
\centering
\includegraphics[width=0.485\textwidth]{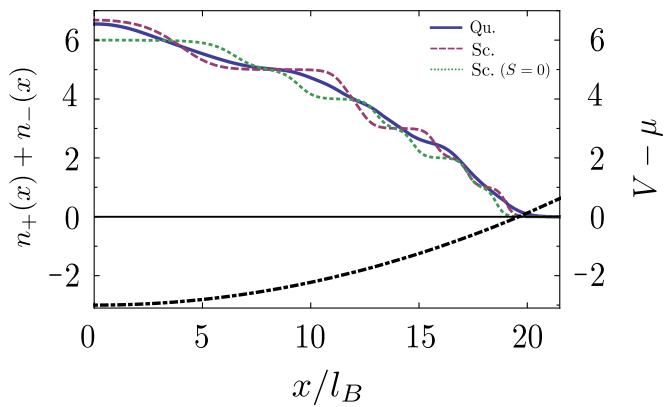} 
\caption{(Color online) Quantum (solid blue curve) and semiclassical (dashed purple and green dotted curves) local electronic density $n(x)$ [in units of $1/(2 \pi l_B^2)$] as a function of the normalized electron position $x/l_B$ for a quadratic 1D potential (shown in units of $\hbar \omega_c$ as a shifted dashed-dotted black half parabola) with $\omega_0 = \omega_c / 8$ and equilibrium chemical potential $\mu = 3 \hbar \omega_c$. In the three cases, relatively high temperature is chosen as $k_B T / (\hbar \omega_c) = 0.06$. Pertinent numerical values for the dressed SO and Zeeman parameters are taken from Ref. \onlinecite{Hernangomez2013}, $S = 0.88$ and $Z= -0.37$, and correspond to typical values experimentally found in InSb semiconductors, characterized by strong SO interaction. In the semiclassical approximation, comparison is made between Eq. \eqref{density_full_scl} in the presence (dashed purple line) and the absence (green dotted line) of Rashba SO interaction. Note the significant difference in the position and width of the density plateaus between the $S\neq 0$ and the $S=0$ cases.
}
\label{fig1}  
\end{figure}

Moving now to the spin polarization, the semiclassical expression valid at high magnetic fields reads
\begin{equation}\label{polarization_scl} 
\Pi_{\textnormal{sc}}^z(\mathbf{r}) = \dfrac{\hbar}{4 \pi l_B^2}\sum_{n=0}^{+\infty} \sum_{\lambda} \lambda n_{\textnormal{F}}[\xi_{n,\lambda}(\mathbf{r})] \sqrt{1 - \dfrac{nS^2}{\Delta^2_n}},
\end{equation}
which shows an explicit dependence on the Rashba SO parameter (again, this is a consequence of the real spin $\sigma$ not being a good quantum number). The expression for the  semiclassical spin polarization for electrons with spin directed towards the $\hat{\mathbf{z}}$ axis agrees with the expectation value of the operator $\hat{s}_z = (\hbar/2) \sigma_z$ calculated in Refs. \onlinecite{Shen2004} and \onlinecite{Shen2005}, except for a global minus sign (which can be attributed to the opposite orientation of the external magnetic field $\mathbf{B}$).

\begin{figure}[tb] 
\centering
\includegraphics[width=0.475\textwidth]{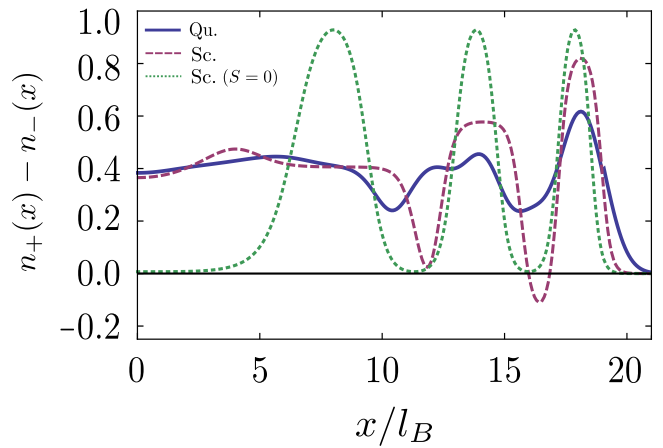} 
\caption{(Color online) Quantum (solid blue curve) and semiclassical (dashed purple and green dotted curves) local spin polarization $\Pi^z(x)$ [in units of $\hbar/(4 \pi l_B^2)$] as a function of the electron position $x/l_B$ for a quadratic 1D potential (not shown here) with $\omega_0 = \omega_c / 8$, equilibrium chemical potential $\mu = 3 \hbar \omega_c$ and relatively high temperature $k_B T / (\hbar \omega_c) = 0.06$. We use the same numerical values for the magnetic field dressed SO and Zeeman parameters as in Fig. \ref{fig1}. For the semiclassical expression \eqref{polarization_scl} both the physical situations $S\neq 0$ (dashed purple line) and $S=0$ (green dotted line) are presented. The spin polarization oscillates within the leading quantum expression \eqref{polarization}, considerably smoothed compared to the semiclassical formula \eqref{polarization_scl}. For strong SO interaction, the semiclassical spin polarization can even change its sign at particular spatial regions.
}
\label{fig2}  
\end{figure}

As an illustration of the obtained functionals, let us consider a toy model given by a 1D quadratic potential $V(x) = (1/2) m^\ast \omega_0^2 x^2$ with characteristic confinement energy $\omega_0 = \omega_c /8$ (so that the potential is smooth enough on the scale of the magnetic length). The energy dispersion relation $\xi_{n,\lambda}(x) = E_{n,\lambda} + V(x)$ consists of a set of energy-shifted parabolas as a function of the electronic coordinate $x$. The resulting spatial dependence of the dimensionless electron density is shown in Fig. \ref{fig1} for not too low temperatures. We present both the results for the quantum formula \eqref{density_per_spin_leading} and for the semiclassical (local) expression \eqref{density_full_scl} that yields the value of the local density after integrating out the quantum fluctuations (note that the relevant behavior is actually provided by the quantum expression at the temperature considered in Fig.  \ref{fig1}). The spatial dependence of the particle density can be seen to be quantized in precise integral steps within the semiclassical approximation, while it is strongly smoothed when taking into account the interference effects of the quantized orbital motion. It is noteworthy that the semiclassical curves present quite different spatial dispersion when comparing the situation of finite SO with zero SO interaction (e.g., different widths and positions of the density plateaus), especially when higher energy levels are involved (i.e., near the center of the potential well) for which the energy spectra are markedly different. However, the differences are strongly reduced at the level of the quantum expressions (the quantum curve for $S=0$ is not shown for the sake of readiability).

In Fig. \ref{fig2} we present the spatial distribution of the dimensionless spin polarization for the same confinement potential model and temperatures that were used in Fig. \ref{fig1}. We also show the results for the quantum formula, given by the combination of Eqs. \eqref{polarization} and \eqref{density_per_spin_leading}, and the semiclassical expression \eqref{polarization_scl}, the latter both in the presence and absence of Rashba SO interaction. Obviously, the primary effect of SO coupling is to reduce the amplitude of the polarization spatial oscillations. The quantum result is also characterized by a smoother spatial dependence than the semiclassical one, as happens with the particle density. However, we note the presence of characteristic beatings for non-zero Rashba SO coupling (both within the quantum and semiclassical results) which can be regarded as a consequence of the different spatial dispersions of each of the spin-projected components of the particle density. Interestingly, the spin polarization can even change its sign in certain spatial regions. This feature, clearly seen with the semiclassical result in Fig. \ref{fig2}, can be understood if we consider the successive filling of the first two Rashba-Landau levels on the basis of Eq. \eqref{polarization_scl}: starting from the edge of the confinement, the first contribution coming from the state $(1,+)$ is positive with a reduced amplitude $\sqrt{1-S^2/\Delta^2_1} $ and is then compensated by the filling of the state $(0,-)$ which has a larger negative amplitude. As a result, the spin polarization becomes negative in some spatial region. Note that this behavior is generic in the sense that it may also occur with the quantum expression (for different values of the parameters, not shown here) and is not limited to the boundaries of the potential. This change in sign is a clear signature of Rashba coupling which could be experimentally studied using spin-polarized local probes to measure the spatial dispersion of the spin imbalance.

To conclude this section, we would like to emphasize the usefulness of the quantum functional  \eqref{density_per_spin_leading} to address electron-electron interaction effects at the  mean-field level in the integer quantum Hall regime. Indeed, the actual electrostatic potential may be very different from the bare electrostatic one (related to confining gates and impurity disorder) due to screening effects, and has to be determined self-consistently from screening theory. Original work\cite{Ch1992,Lier1994} investigating screening properties at high magnetic fields used semiclassical expressions such as formula \eqref{density_full_scl} as a starting point for the calculations, and got the physical picture of the formation in the sample of alternating compressible and incompressible regions of different widths. Further work\cite{Siddiki2004} with simple 1D toy models in the Hartree approximation has shown that quantum smearing effects lead to important quantitative deviations in the pattern of these regions, thus justifying the need to work preferentially with general quantum functionals such as \eqref{density_per_spin_leading} for realistic calculations.

\section{Electron Current Density}\label{sec_electron_current}

\subsection{General expression}
In this second half of the paper, we are interested in the electron current density.
We first consider the electron current density operator $\hat{\mathbf{j}}$ in the presence of Rashba SO interaction given by the general formula
\begin{equation}
 \hat{\mathbf{j}} = e \dfrac{i}{\hbar} [\hat{\mathcal{H}} ,\hat{\mathbf{r}}],
\end{equation}
where the Hamiltonian is defined in Eq. \eqref{Hamiltonian}. Evaluation of the commutator gives an expression for the local electron current density operator which can be written as the sum of two components $ \hat{\mathbf{j}}= \hat{\mathbf{j}}_1+ \hat{\mathbf{j}}_2$, the first being the usual U$(1)$ electron current density operator diagonal in spin space
\begin{equation}
 \hat{\mathbf{j}}_1= \dfrac{e}{m^\ast} \left[\hat{\mathbf{p}} - \dfrac{e}{c} \mathbf{A}(\hat{\mathbf{r}}) \right] \otimes 1\!\!1_s,
\end{equation}
and the second a spatially constant term
\begin{equation}
 \hat{\mathbf{j}}_2 = \dfrac{e^2}{m^\ast c} \tilde{\mathbf{A}},
\end{equation}
proportional to a fictitious SO-dependent vector potential 
\begin{equation}
 \tilde{\mathbf{A}} = -\dfrac{\alpha m^\ast c}{e} \hat{\mathbf{z}} \times \bm{\sigma}.
\end{equation}
The fictitious vector potential $\tilde{\mathbf{A}}$ can be understood as a non-Abelian SU$(2)$ gauge field with the Rashba SO parameter playing the role of an SU$(2)$ coupling constant [analogous to the electric charge in the U$(1)$ formulation of electrodynamics]. This point of view has been used in the literature\cite{Tokatly2008, Vignale2012} since, in principle, it ensures proper definitions of spin-related quantities such as the spin current from covariant conservation laws.

Using the local equilibrium distribution function \eqref{Glesserdefinition} written in the electronic representation, we get the expression of the spectral (electron) current density by a trace of the former with the current density operator in which the canonical momentum operator has been substituted by its representation in real space $\hat{\mathbf{p}}\rightarrow -i\hbar \bm{\nabla}_{\mathbf{r}}$: 
\begin{equation}
\mathbf{j}(\mathbf{r},\omega) = \textnormal{Tr}\left\{{\mathbf{j}}({\mathbf{r},\mathbf{r}'})[-iG^{<}(\mathbf{r},\mathbf{r}', \omega)]\right\}\Big|_{\mathbf{r} = \mathbf{r}'},
\end{equation}
where ${\mathbf{j}}({\mathbf{r},\mathbf{r}'})=[ \langle {\bf r} |\hat{\mathbf{j}} | {\bf r} \rangle +\langle {\bf r}' |\hat{\mathbf{j}} | {\bf r}' \rangle ^{\ast}]/2$. 
Here, as we did before with the electron density, we choose to express the spectral current density as a sum over two spin-resolved components
\begin{equation}
\mathbf{j}(\mathbf{r},\omega) = \sum_{\sigma=\pm} \mathbf{j}_{\sigma}(\mathbf{r},\omega).
\end{equation}
It is convenient to write both the U$(1)$ and SU$(2)$ components of the current in terms of the electronic kernels and the vortex spectral density in the following way  
\begin{align}\label{electron_density_1_general}
\mathbf{j}_{1\sigma}(\mathbf{r},\omega) &= \dfrac{e\hbar}{2m^\ast} \int \dfrac{d^2 \mathbf{R}}{2 \pi l_B^2} \sum_{n_1,\lambda_1} \sum_{n_2,\lambda_2} f_{\sigma}(\theta_{n_1}^{\lambda_1})f_{\sigma}(\theta_{n_2}^{\lambda_2}) \notag \\ &\times \Big[(\bm{\nabla}_{\mathbf{r}'}-\bm{\nabla}_{\mathbf{r}}) + \dfrac{2ie}{\hbar c} \mathbf{A}(\mathbf{r})\Big] K_{n_{1\sigma};n_{2\sigma}}(\mathbf{r},\mathbf{r}',\mathbf{R}) \notag \\ &\times g^{<}_{n_1,\lambda_1;n_2,\lambda_2}(\mathbf{R},\omega)\Bigg|_{\mathbf{r} = \mathbf{r}'},
\end{align}
\begin{align}\label{electron_density_2_general}
& \mathbf{j}_{2\sigma}(\mathbf{r},\omega) = \alpha e \int \dfrac{d^2\mathbf{R}}{2 \pi l_B^2}\sum_{n_1,\lambda_1}\sum_{n_2,\lambda_2} f_{\sigma}(\theta_{n_1}^{\lambda_1})f_{-\sigma}(\theta_{n_2}^{\lambda_2}) \notag \\ &\times K_{n_{1\sigma};n_{2-\sigma}}(\mathbf{r},\mathbf{r},\mathbf{R})g^{<}_{n_1,\lambda_1;n_2,\lambda_2}(\mathbf{R},\omega)
\begin{pmatrix}
\sigma \\
i
\end{pmatrix}
.
\end{align}
Equation \eqref{electron_density_1_general} can be further manipulated following the steps detailed in Ref. \onlinecite{Champel2008} for the spinless case to get the formula
\begin{multline}\label{electron_density_1_general_simplified}
\mathbf{j}_{1\sigma}(\mathbf{r},\omega)=-\dfrac{ie\hbar}{2m^\ast}  \int \dfrac{d^2 \mathbf{R}}{2 \pi l_B^2} \sum_{n_1,\lambda_1} \sum_{n_2,\lambda_2} f_{\sigma}(\theta_{n_1}^{\lambda_1})f_{\sigma}(\theta_{n_2}^{\lambda_2}) \\ \times \mathbf{J}_{n_{1\sigma};n_{2\sigma}}(\mathbf{r},\mathbf{R})\,g^{<}_{n_1,\lambda_1;n_2,\lambda_2}(\mathbf{R},\omega),
\end{multline}
with $\mathbf{J}_{n_{1\sigma},n_{2\sigma}}(\mathbf{r},\mathbf{R})$ an electronic current kernel that contains the contribution of the orbital motion and is defined as
\begin{align}\label{current_kernel}
 \mathbf{J}_{n_{1\sigma};n_{2\sigma}}&(\mathbf{r},\mathbf{R})=\hat{\mathbf{z}} \times \bm{\nabla}_\mathbf{r}K_{n_{1\sigma};n_{2\sigma}}(\mathbf{r},\mathbf{r},\mathbf{R})  - \dfrac{i\sqrt{2}}{l_B} \notag \\ &\times \Bigg[ \sqrt{n_{2\sigma}+1} K_{n_{1\sigma};n_{2\sigma}+1}(\mathbf{r},\mathbf{r},\mathbf{R})\begin{pmatrix}
 1\\
 i                                                                                                                                                                                                                                                                                                                                                                                                                                                                                                                                                                                                      \end{pmatrix} 
 \notag \\ &+ \sqrt{n_{1\sigma}+1}K_{n_{1\sigma}+1;n_{2\sigma}}(\mathbf{r},\mathbf{r},\mathbf{R})\begin{pmatrix}
 -1\\
 i                                                                                                                                                                                                                                                                                                                                                                                                                                                                                                                                                                                                      \end{pmatrix}\Bigg].
\end{align}

These equations are completely general and allow us, in principle, to obtain the electron current density at arbitrary (finite) order in the $l_B$ expansion collecting the contributions that arise from the potential matrix elements and the SO vortex Green's function. 
In the present article, we are only interested in the most significant contribution to the electron current density at high magnetic fields.
Note that this contribution actually comes both from the dominant and subdominant terms in the SO vortex Green's function expansion\cite{Champel2008}, because the current density contains the large prefactor $\hbar \omega_c$ [in contrast, the knowledge of the leading order vortex Green's function $g^{(0)}(\mathbf{R},\omega)$ was sufficient to fully determine the electron density at high magnetic fields]. In the following we will distinguish these two terms in the current density which originate from $g^{(0)}(\mathbf{R},\omega)$ and $g^{(1)}(\mathbf{R},\omega)$ and correspond to the so-called density-gradient and drift current densities, respectively. 

\subsection{Density-gradient current}
Since the leading-order SO vortex Green's function is diagonal in the Rashba-Landau and SO quantum numbers, the current kernel for the U$(1)$ electron current density in Eq. \eqref{current_kernel} will present combinations of electronic kernels of the form $\sqrt{n+1} K_{n;n+1}(\mathbf{r},\mathbf{r},\mathbf{R}) \pm \textnormal{c.c.}$ (where $ \textnormal{c.c.}$ stands for complex conjugate). Each term satisfies the following useful identity\cite{Champel2008}
\begin{equation}\label{useful_identity}
 \sqrt{n+1} K_{n;n+1}(\mathbf{r},\mathbf{r},\mathbf{R})= \dfrac{l_B}{\sqrt{2}} \sum_{q=0}^{n}\left[\partial_{X}-i\partial_{Y} \right]K_q(\mathbf{r}-\mathbf{R}).
\end{equation}
so that the current kernel is reduced to
\begin{equation}
\mathbf{J}_{n_\sigma;n_\sigma}(\mathbf{r},\mathbf{R}) = \hat{\mathbf{z}} \times \bm{\nabla}_{\mathbf{r}} \Big[K_{n}(\mathbf{r}-\mathbf{R})-2 \sum_{q=0}^{n_\sigma}K_{q}(\mathbf{r}-\mathbf{R}) \Big],
\end{equation}
after having used the relation $\bm{\nabla}_\mathbf{R} K_n(\mathbf{r}-\mathbf{R}) = -\bm{\nabla}_\mathbf{r} K_n(\mathbf{r}-\mathbf{R})$. Note that the vector product coming from the current kernel ensures that this contribution to the current density has zero bulk average.

Inserting Eq. \eqref{Green_zero_lesser} into Eq. \eqref{electron_density_1_general_simplified} we perform an integration over the energies [i.e., we define $\mathbf{j}(\mathbf{r})=(2\pi)^{-1} \int d\omega\, \mathbf{j}(\mathbf{r},\omega)$] to obtain the following quantum expression for the U$(1)$ density current:
\begin{multline}
 \mathbf{j}^{\textnormal{dg}}_{1\sigma}(\mathbf{r})= -\dfrac{e}{h}(\hbar \omega_c)\, \hat{\mathbf{z}} \times \bm{\nabla}_\mathbf{r} \int d^2 \mathbf{R} \sum_{n=0}^{+\infty}\sum_{\lambda} f_{\sigma}^2(\theta_n^\lambda)  \\ \times n_{\textnormal{F}}[\xi_{n,\lambda}(\mathbf{R})]  
 \left[\sum_{q=0}^{n_\sigma}K_{q}(\mathbf{r}-\mathbf{R}) - \dfrac{1}{2}K_{n_\sigma}(\mathbf{r}-\mathbf{R})\right]. \label{j1quantum}
\end{multline}
For the SU$(2)$ part of the electron current density we follow analogous steps: we first insert the leading-order lesser SO vortex Green's function \eqref{Green_zero_lesser} into Eq. \eqref{electron_density_2_general} and then perform an integration over the energies to get
\begin{multline}
 \mathbf{j}^{\textnormal{dg}}_{2 \sigma}(\mathbf{r})= i \alpha e \int \dfrac{d^2 \mathbf{R}}{2 \pi l_B^2} \sum_{n=1}^{+\infty} \sum_{\lambda} f_{\sigma}(\theta_n^\lambda) f_{-\sigma}(\theta_n^\lambda) \\ \times n_{\textnormal{F}}[\xi_{n,\lambda}(\mathbf{R})] K_{n_\sigma;n_{-\sigma}}(\mathbf{r},\mathbf{r},\mathbf{R}) 
 \begin{pmatrix}
\sigma \\
i
\end{pmatrix}.
\end{multline}
Now, we shall only consider  the quantity ${\bf j}^{\textnormal{dg}}_2({\bf r})$ resulting from the sum over the two spin-projected contributions. The technical reason for this is that we need to combine terms coming from the off-diagonal elements of the electronic kernels $ K_{n;n+1}(\mathbf{r},\mathbf{r},\mathbf{R})$ with their complex conjugates to obtain a real  expression for the total current  density. Again, this can be physically understood through the observation that in the presence of SO interaction, the spin $\sigma$ is not a good quantum number and there is \textit{a priori} no reason why the electron current density can be split in terms of two independent spin-resolved components. Performing the sum over $\sigma$ and using Eq. \eqref{useful_identity} we obtain
\begin{multline}
\mathbf{j}^{\textnormal{dg}}_2(\mathbf{r})= \dfrac{e}{h}(\hbar \omega_c)\dfrac{S}{2}\hat{\mathbf{z}} \times \bm{\nabla}_{\mathbf{r}} \int d^2 \mathbf{R} \sum_{n=1}^{+\infty} \sum_{\lambda}  f_{+}(\theta_n^\lambda)f_{-}(\theta_n^\lambda) \\ \times  \dfrac{1}{\sqrt{n}} n_{\textnormal{F}}[\xi_{n,\lambda}(\mathbf{R})] \sum_{q=0}^{n-1} K_q(\mathbf{r}-\mathbf{R}). \label{j2quantum}
\end{multline}

As we did previously with the local electron density, we can consider the semiclassical limit within the current density quantum functionals \eqref{j1quantum} and \eqref{j2quantum}. This yields the semiclassical result for the total current density: 
\begin{multline}\label{density_gradient_semiclassical}
\mathbf{j}^{\textnormal{dg}}_{\textnormal{sc}}(\mathbf{r})= \dfrac{e}{h}(\hbar \omega_c) \sum_{n=0}^{+\infty}\sum_{\lambda}\sum_{\sigma=\pm} \dfrac{1}{2}\Bigg\{\left(n_\sigma + \dfrac{1}{2} \right) \\ 
\times \left[ 1 -\sigma \lambda \sqrt{1-\dfrac{n S^2}{\Delta^2_n}}\right] - \dfrac{\lambda}{4}\dfrac{nS^2}{\Delta^2_n}\Bigg\}
 \bm{\nabla}_{\mathbf{r}} n_{\textnormal{F}}[\xi_{n,\lambda}(\mathbf{r})] \times \hat{\mathbf{z}}.
\end{multline}
Expression \eqref{density_gradient_semiclassical} corresponds to the so-called density-gradient\cite{Champel2008} (or {edge}\cite{Vignale1994}) contribution to the electronic current density since it represents an electron flow appearing as a response to a change in the particle density. In this formula we can see that the  dependence on the dressed Rashba SO parameter has two different origins: on one hand, the weights of the SO vortex wave functions [they appear both in the U(1) and the SU(2) contributions]; on the other hand, the particular functional form of the SU(2) current density operator which introduces a term linear in $S$ into the second contribution [see Eq. \eqref{j2quantum}]. In principle, both U$(1)$ and SU$(2)$ contributions can be discriminated in the high-temperature regime by their different dependencies on both the Rashba-Landau level index and the magnetic field. The clear signatures of each of the terms suggest future experiments to experimentally obtain the values of the Rashba SO parameter $\alpha$ in 2DEGs by performing a careful analysis of the spatial structure of the edge states. 

\subsection{Drift current}

The drift (or {bulk}\cite{Vignale1994}) contribution to the electronic current density appears\cite{Champel2008} when considering the first order (subdominant) correction to the leading-order (diagonal) SO vortex Green's function, and characterizes Rashba-Landau level mixing processes related to gradients of the potential energy. To compute this drift term, we first need the subleading (off-diagonal) component of the distribution function which can be obtained by inserting the SO vortex Green's function given in  Eq. \eqref{Green_SO_first} into Eq. \eqref{Glesserdefinition}. Next, using the general expressions for the contributions to the  current density written in Eqs. \eqref{electron_density_2_general} and \eqref{electron_density_1_general_simplified} we obtain the spin-resolved U$(1)$ term after some algebra,
\begin{align}\label{current_U1}
 \mathbf{j}_{1\sigma}^{\textnormal{dr}}&(\mathbf{r})= \dfrac{e}{h}\int d^2\mathbf{R} \sum_{n=0}^{+\infty} \sum_{\lambda_1,\lambda_2} \sum_{\sigma'=\pm} \sqrt{\dfrac{n_{\sigma'}+1}{n_\sigma+1}} \notag \\ &\times  f_\sigma(\theta_{n}^{\lambda_1})f_{\sigma'}(\theta_{n}^{\lambda_1})f_\sigma(\theta_{n+1}^{\lambda_2})f_{\sigma'}(\theta_{n+1}^{\lambda_2}) F_{n;\lambda_1;\lambda_2}(\mathbf{R}) \, \hat{\mathbf{z}} \notag \\ & \times \left[ \bm{\nabla}_\mathbf{R}V(\mathbf{R}) \sum_{q=0}^{n_\sigma}K_q(\mathbf{r}-\mathbf{R}) \right.  \notag \\ & \left. +\,l_B^2    \left[\bm{\nabla}_\mathbf{R}V(\mathbf{R}) \cdot  \bm{\nabla}_\mathbf{r} \right] \bm{\nabla}_\mathbf{r} \sum_{q=0}^{n_\sigma}(n_{\sigma}+\frac{1}{2}-q) K_q(\mathbf{r}-\mathbf{R}) \right],
 \end{align}
while the full SU$(2)$ term reads
\begin{align}\label{current_U2}
 \mathbf{j}^{\textnormal{dr}}_{2}&(\mathbf{r})=-\dfrac{e}{h} \dfrac{S}{2} \int d^2 \mathbf{R} \sum_{n=0}^{+\infty}\sum_{\lambda_1,\lambda_2} \sum_{\sigma'=\pm}\sqrt{n_{\sigma'}+1} \notag\\ \notag &\times f_{\sigma'}(\theta_{n}^{\lambda_1})  f_{\sigma'}(\theta_{n+1}^{\lambda_2}) f_{-}(\theta_n^{\lambda_1}) f_{+}(\theta_{n+1}^{\lambda_2})F_{n;\lambda_1;\lambda_2}(\mathbf{R})\, \hat{\mathbf{z}} \\ &\times  \Bigg( \bm{\nabla}_\mathbf{R} V(\mathbf{R}) \Bigg\{K_n(\mathbf{r}-\mathbf{R}) + \dfrac{1}{\sqrt{n+1}\sqrt{n+2}} \notag \\ &\times \Bigg[\sum_{q=0}^n K_{q}(\mathbf{r}-\mathbf{R})-(n+1)K_{n+1}(\mathbf{r}-\mathbf{R})\Bigg]\Bigg\}  \notag\\ 
 &+\dfrac{l_B^2\left[\bm{\nabla}_\mathbf{R}V(\mathbf{R}) \cdot  \bm{\nabla}_\mathbf{r} \right] \bm{\nabla}_\mathbf{r}}{\sqrt{n+1}\sqrt{n+2}} \sum_{q=0}^{n}(n+1-q) K_q(\mathbf{r}-\mathbf{R})     
 \Bigg).
\end{align}
Here, $F_{n;\lambda_1;\lambda_2}(\mathbf{R})$ denotes a particular combination of Fermi-Dirac distribution functions evaluated for adjacent Rashba-Landau energy levels
\begin{equation}\label{Gamma_function}
F_{n;\lambda_1;\lambda_2}(\mathbf{R})= \hbar \omega_c  \dfrac{ n_{\textnormal{F}}[\xi_{n+1,\lambda_2}(\mathbf{R})]-n_{\textnormal{F}}[\xi_{n,\lambda_1}(\mathbf{R})]
}{E_{n+1,\lambda_2}-E_{n,\lambda_1}}.
\end{equation}
A detailed derivation of both quantum expressions \eqref{current_U1} and \eqref{current_U2} is provided in the Appendix.

These expressions for the drift current density are greatly simplified in the semiclassical limit $l_B \to 0$. For reasons of clarity, we shall thus analyze in the rest of the paper the electronic current for this approximation only with the effects of wave function spreading postponed for future work.
Adding both U(1) and SU(2) contributions and considering the high-temperature regime where we can make the replacement $K_q({\bf r}-{\bf R}) \simeq \delta^{(2)}({\bf r}-{\bf R})$ in Eqs. \eqref{current_U1} and \eqref{current_U2}, we obtain the final expression for the total semiclassical drift current density in the presence of Rashba SO and Zeeman interactions:
\begin{align}\label{drift_current_semiclassical}
&\mathbf{j}_{\textnormal{sc}}^{\textnormal{dr}}(\mathbf{r}) = \dfrac{e}{h} \sum_{n=0}^{+\infty}\sum_{\lambda_1,\lambda_2}\Bigg[\sum_{\sigma'=\pm}\sqrt{n_{\sigma'}+1}f_{\sigma'}(\theta_n^{\lambda_1})f_{\sigma'}(\theta_{n+1}^{\lambda_2}) \Bigg] 
 \notag \\ &\times\!\! \Bigg[\sum_{\sigma=\pm}\sqrt{n_{\sigma}+1} f_{\sigma}(\theta_n^\lambda)f_{\sigma}(\theta_{n+1}^{\lambda_2})  \notag  -\dfrac{S}{2}f_{-}(\theta_n^{\lambda_1})f_{+}(\theta_{n+1}^{\lambda_2}) \Bigg] \\ &\times F_{n;\lambda_1;\lambda_2}(\mathbf{r})\, \hat{\mathbf{z}} \times \bm{\nabla}_\mathbf{r} V(\mathbf{r}).
\end{align}

\begin{figure} 
\centering
\includegraphics[width=0.48\textwidth]{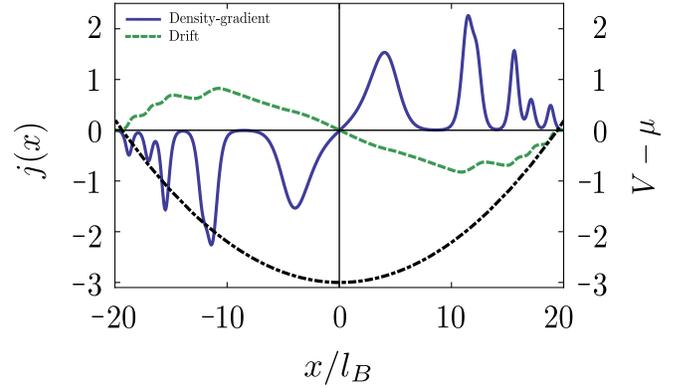} 
\caption{(Color online) Semiclassical density-gradient \eqref{density_gradient_semiclassical} and drift \eqref{drift_current_semiclassical} current density contributions [in units of $e \omega_c / (2\pi l_B)$] as a function of the normalized electron position $x/l_B$ for a quadratic 1D potential (represented in units of $\hbar \omega_c$ by the shifted dashed-dotted parabola) with $\omega_0 = \omega_c / 8$, $\mu = 3 \hbar \omega_c$ and $k_B T / (\hbar \omega_c) = 0.06$. Here, the SO and Zeeman parameters $S$ and $Z$ are chosen to be the same as in Fig. \ref{fig1}. The broadened peaks representing the edge states in the density-gradient component of the local current density show the spatial structure of the energy spectrum. Note the alternating sign of each of the contributions that combine to give non-negligible spatial oscillations of the total current density.}
\label{fig3}  
\end{figure}

Figure \ref{fig3} displays the spatial distribution of the density-gradient and drift components of the equilibrium current density on the basis of semiclassical Eqs. \eqref{density_gradient_semiclassical} and \eqref{drift_current_semiclassical} for the quadratic 1D potential $V(x) = (1/2)m^\ast \omega_0^2 x^2$ with $\omega_0 = \omega_c / 8$. The density-gradient contribution presents sharp peaks broadened by temperature whose spatial structure can be used to determine the spectral dispersion of the edge states. Indeed, the minima of the density-gradient current can be perfectly matched with the electron density plateaus shown in Fig. \ref{fig1}.  On the other hand, the drift contribution is a very smooth curve showing very little sign of the presence of each of the edge states. At $T=0$, the density-gradient current flow can be seen to vanish in the incompressible regions of the disordered 2DEG characterized by constant particle density while being nonzero in the compressible (inhomogeneous) parts. For $T > 0$ the combination of both terms (with different sign) produces spatial oscillations in the total current density that can be attributed to the fact that the compressible stripes and the regions of strong potential gradients yield opposite currents at finite temperature.\cite{Vignale1994}

To conclude this section, we may mention that it is not difficult to prove that the well-known expression for the drift current density can be smoothly recovered in the limit of vanishing SO coupling ($|S| \rightarrow 0$). Indeed, in this limit the angular parameters $\theta_{n}^{\lambda} \rightarrow (1+ \lambda)\pi/4$; i.e.,  $f_{\pm}(\theta_n^{\pm})=1$ independently of the Rashba-Landau level index. The SU$(2)$ contribution to the semiclassical drift current density vanishes trivially and we obtain from the U$(1)$ part
\begin{multline}
\mathbf{j}^{\textnormal{dr}}_{\textnormal{sc}}(\mathbf{r})_{\alpha\rightarrow 0} = \dfrac{e}{h} \sum_{n=0}^{+\infty} \sum_{\sigma=\pm}  n_{\textnormal{F}}[\xi_{n,\sigma}(\mathbf{r})]\bm{\nabla}_{\mathbf{r}} V(\mathbf{r}) \times \hat{\mathbf{z}},
\end{multline}
where the Zeeman-split effective energies in the presence of an electrostatic potential are given by the expression  $\xi_{n,\sigma}(\mathbf{r}) = \hbar \omega_c(n+1/2+ \sigma Z/2) + V(\mathbf{r})$. 

\section{Transport properties}\label{subsec_nonequilibrium}

\subsection{Local Hall conductivity}

As mentioned previously, a great advantage of our high magnetic field Green's function theory is the possibility to derive microscopic expressions for different local observables under the form of functionals of the potential energy $V(\mathbf{r})$. This general character somehow allows us to easily deal with the additional presence of nonequilibrium electric fields, as occurring in a transport experiment. In the nonperturbative high magnetic field regime, equilibrium and nonequilibrium local electric fields have in fact to be treated on an equal footing, and we can directly use the results of Sec. \ref{sec_electron_current} to derive a nonequilibrium current density.
At the linear response level, we can consider that the nonequilibrium electrostatic potential has the simple form $V_\textnormal{neq}(\mathbf{r})= e \Phi(\mathbf{r})$, corresponding to the electromotive field $\mathbf{E}(\mathbf{r})= -\nabla_{\mathbf{r}}\Phi(\mathbf{r})$. This electromotive field gives rise to a nonequilibrium drift current density whose expression, at relatively high temperatures (quantum coherence effects being neglected), can be directly inferred from semiclassical formula \eqref{drift_current_semiclassical}. Not surprisingly, the drift current density in this hydrodynamic high magnetic field regime adopts the form of a local Ohm's law $\mathbf{j}_{\textnormal{neq}}(\mathbf{r}) =\bar{\sigma}(\mathbf{r}) \mathbf{E}(\mathbf{r}) = \sigma_\textnormal{H}(\mathbf{r}) \hat{\mathbf{z}} \times \mathbf{E}(\mathbf{r})$, where the purely off-diagonal conductivity tensor $\bar{\sigma}(\mathbf{r})$ has a Hall component
\begin{align}\label{Hall_conductivity_full}
&\sigma_{\textnormal{H}}(\mathbf{r})=-\dfrac{e^2}{h} \sum_{n=0}^{+\infty}\sum_{\lambda_1,\lambda_2}\Bigg[\sum_{\sigma'=\pm}\sqrt{n_{\sigma'}+1}f_{\sigma'}(\theta_n^{\lambda_1})f_{\sigma'}(\theta_{n+1}^{\lambda_2}) \Bigg]
 \notag \\ &\,\,\,\times  \Bigg[\sum_{\sigma=\pm}\sqrt{n_{\sigma}+1} f_{\sigma}(\theta_n^\lambda)f_{\sigma}(\theta_{n+1}^{\lambda_2})     -\dfrac{S}{2}f_{-}(\theta_n^{\lambda_1})f_{+}(\theta_{n+1}^{\lambda_2}) \Bigg] \notag \\ & \,\,\, \times \hbar \omega_c \dfrac{n_{\textnormal{F}}[\xi_{n+1,\lambda_2}(\mathbf{r})] - n_{\textnormal{F}}[\xi_{n,\lambda_1}(\mathbf{r})]}{E_{n+1,\lambda_2}-E_{n,\lambda_1}}.
\end{align}
Equation \eqref{Hall_conductivity_full} constitutes one of the main results of this article. It describes the local semiclassical Hall conductivity in the presence of both uniform Rashba SO and Zeeman couplings when the electron is placed in a smooth disorder potential. The resulting expression is highly nontrivial (dependence on two SO quantum numbers $\lambda_1$ and $\lambda_2$ simultaneously with mixing between adjacent Rashba-Landau levels indices) and brings important information on the characteristic values of the Hall plateaus depending on the strength of the Rashba SO interaction.

As a first impression, one may naively foresee that the derived expression \eqref{Hall_conductivity_full} does not lead to a quantization of the Hall conductivity in units of the conductance quantum, $e^2/h$, at low temperatures due to the Rashba SO interaction. Notwithstanding, it can be shown by symbolic calculation that Eq. \eqref{Hall_conductivity_full} can be simplified for any {\em finite} value of the SO parameter $S$ into
\begin{equation}\label{Hall_conductivity_simpl}
\sigma_{\textnormal{H}}(\mathbf{r})= \dfrac{e^2}{h}\sum_{n=0}^{+\infty} \sum_{\lambda} n_{\textnormal{F}}[\xi_{n,\lambda}(\mathbf{r})].
\end{equation}
This result was already suggested by the semiclassical expression \eqref{density_full_scl} for the electron density. Indeed, at the semiclassical level we can use the classical Hall formula that relates the local electron density and the Hall conductivity $\sigma_\textnormal{H}(\mathbf{r}) = (ec/B)n(\mathbf{r})$ to directly obtain Eq. \eqref{Hall_conductivity_simpl} (this is, of course, not true from the moment that we use the full quantum expression for the electron current density that includes the electronic kernels). We thus conclude that the local semiclassical Hall conductivity is robust with respect to any {\em finite} (but not infinite) Rashba SO coupling and quantized in units of $e^2 / h$ at low temperatures. 

\subsection{Half-integer quantization of the Hall conductivity in a disordered topological insulator} 
Our general formula  \eqref{Hall_conductivity_full} can also be used to study the local Hall conductivity on (the surface of) a topological insulator. For low energies, the latter can be described by a simple model consisting in a single Dirac cone, which can be obtained from an effective Hamiltonian for a finite film within the slab geometry neglecting any possible coupling between the two (top and bottom) surfaces and all the contributions from the bulk nonlinear in the momentum. Here, this situation can be straightforwardly achieved in the disordered 2DEG with Rashba SO interaction and Zeeman coupling by simply taking the limit $|S| \rightarrow +\infty$ in the local conductivity expression \eqref{Hall_conductivity_full} (this corresponds to the simultaneous formal limits $m^\ast \rightarrow +\infty$ and $g\rightarrow 0$). 

First, for the energy levels \eqref{spectrum}, this limit yields a gapless and purely relativistic-like energy spectrum
\begin{equation}\label{spectrum_linear}  
\tilde{\xi}_{n,\lambda}(\mathbf{r}) = - \lambda \sqrt{n} \hbar \Omega + V(\mathbf{r}),  
\end{equation} 
where $\Omega = \alpha \sqrt{2} / l_B$ is a SO-dependent characteristic frequency (we note that this frequency has the same form as the characteristic frequency of graphene with the SO Rashba parameter playing the role of a spatially constant Fermi velocity). As happens with effective Dirac Hamiltonians describing relativistic-like electrons in a perpendicular magnetic field, the characteristic energy is no longer proportional to the magnetic field, $\hbar \omega_c \propto B$, but $\hbar \Omega \propto \sqrt{B}$. Very importantly, we note that the spectrum \eqref{spectrum_linear} is no more bounded from below, as a difference to Eq. \eqref{spectrum} which was for any {\em finite} values of $S$. This usually imposes to soundly alter the conductivity formula, since there is now an infinite number of filled Rashba-Landau levels below the chemical potential $\mu$. This situation takes also place in graphene where the density (and consequently) the Hall 
conductivity have to be defined rather in terms of the population imbalance between electrons and holes. 

We now show that the correct result for the Hall conductivity can be simply obtained in our scheme, without having to introduce a new definition for the current density (note that other regularization procedures can be used to get the same result\cite{Konig2013}).
In the limit $|S| \to+ \infty$, the angular  parameters $\theta_n^{\lambda}$ appearing in the weighting functions $f_\sigma(\theta_n^{\lambda})$ approach the finite values $\theta_n^{\lambda}= \lambda \pi/4$ independently of the Rashba-Landau level index $n$. Inspection of Eqs. \eqref{Gamma_function} and \eqref{Hall_conductivity_full} then shows that the U$(1)$ part of the expression for the local Hall conductivity tends to zero while the  SU$(2)$ term tends towards a finite value. The Hall conductivity can  be further simplified quite straightforwardly to obtain
\begin{equation}
\sigma_{\textnormal{H}}(\mathbf{r})=   \dfrac{e^2}{h} \sum_{n=0}^{+\infty}\sum_{\lambda} \!\left(n + \dfrac{1}{2} \right) \! \left\{ n_{\textnormal{F}}[\tilde{\xi}_{n,\lambda}(\mathbf{r})] - n_{\textnormal{F}}[\tilde{\xi}_{n+1,\lambda}(\mathbf{r})]\right\}, \label{HallS}
\end{equation} 
which is the well-known result for the Hall conductivity in low-energy graphene\cite{Gusynin2005}, up to a global prefactor (Hall conductivity in graphene appears to be four times bigger due to the presence of spin and valley degeneracies). 
Consequently, we expect a sequence of Hall plateaus pinned at  values\cite{footnote4} $\pm 1/2$, $\pm 3/2$, $ \pm 5/2$,\dots of the conductance quantum. Formula \eqref{HallS} can be (only formally) rewritten as
\begin{equation} 
 \sigma_\textnormal{H}(\mathbf{r}) =  \dfrac{e^2}{h} \left\{ \dfrac{1}{2}n_{\textnormal{F}}[\tilde{\xi}_{0,-}(\mathbf{r})] +\sum_{n=1}^{+\infty} \sum_{\lambda=\pm} n_{\textnormal{F}}[\tilde{\xi}_{n,\lambda}(\mathbf{r})]\right\}, 
\end{equation} 
which clearly shows that the lowest Rashba-Landau energy level $n=0$ presents half of the degeneracy in comparison with higher Rashba-Landau levels. This interesting result had not been obtained by a microscopic theory in the presence of disorder until now, although standard calculations using the Kubo formula for the clean system in the presence of surface hybridization exist in the literature.\cite{Burkov2011} Half-integer quantization of the lowest Landau level in Dirac massless Hamiltonians without disorder was already predicted three decades ago and is linked to the fact that for $n=0$, we always have a bound state with zero energy due to the supersymmetric structure of the Hamiltonian.\cite{Jackiw1984} 

\subsection{Spin Hall conductivity in the quantum Hall regime} \label{spin_polarization_sec} 

We have previously noted in Sec. \ref{sec_electron_density} that it is possible to define at high magnetic fields a dressed spin quantum number $\lambda=\pm$, which takes into account the effect of SO Rashba coupling. This $\lambda$ remains a good quantum number in the presence of a smooth disordered potential, as exemplified by the semiclassical local Hall conductivity \eqref{Hall_conductivity_simpl} which is written as a sum of two {\em independent} $\lambda$-resolved  components. In fact, this can already be established from the quantum expressions \eqref{current_U1} and \eqref{current_U2} for the total drift current density, which can be generally decomposed as
\begin{eqnarray}
{\bf j}({\bf r})=\sum_{\lambda} {\bf j}_{\lambda}({\bf r}).
\end{eqnarray}
From the physical point of view, this property can be understood as a consequence of the installation of a local equilibrium regime in the 2DEG at high magnetic fields. At leading order of the $l_B$ expansion ($l_B=0$), the angle $\theta_n^\lambda$ related to the Rashba precession in spin space is independent of the local electrostatic potential, and thus remains spatially constant. 
More generally, when  taking into account the quantum  effects ($l_B \neq 0$) for the vortex degree of freedom, the observables can still be expressed\cite{Hernangomez2013} as a sum over two SO-resolved components (with respect to some redefined SO quantum number), but the precession angle then slowly varies with the vortex position (in an adiabatic way), as a result of the smooth local spatial variations of the electrostatic disordered potential.
 
Now, in the semiclassical limit ($l_B \to 0$), the $\lambda$ components of the current density simply read
\begin{equation}\label{SOresolved_current} 
\mathbf{j}_\lambda(\mathbf{r})=\dfrac{e^2}{h}\sum_{n}  n_{\textnormal{F}}[\xi_{n,\lambda}(\mathbf{r})]\, \hat{\mathbf{z}}\times \mathbf{E}(\mathbf{r}). 
\end{equation} 
Using this simplified formula \eqref{SOresolved_current} for the SO-polarized electron current, it is also quite easy to obtain the local spin Hall conductivity, $\sigma_\textnormal{H}^s(\mathbf{r})$, for electrons with their spin polarized along the SO-dependent axis for which $\lambda=\pm$ is a good quantum number. 
The associated nonequilibrium local  spin current $\mathbf{j}^s_{\textnormal{neq}}(\mathbf{r})$ can then be interpreted as being a combination of two independent SO-polarized currents in which electrons with different SO quantum number move in opposite directions without net charge flow. With this idea in mind, we write the flow of angular momentum as 
\begin{equation}
\mathbf{j}^s_{\textnormal{neq}}(\mathbf{r})= \dfrac{\hbar}{2e}\Big[ \mathbf{j}_{+}(\mathbf{r})- \mathbf{j}_{-}(\mathbf{r})\Big], 
\end{equation} 
which gives the spin Hall conductivity  
\begin{equation}\label{spinHall_conductivity}  
\sigma_{\textnormal{H}}^s(\mathbf{r})= \dfrac{e}{4\pi} \sum_{n} \Big\{n_{\textnormal{F}}[\xi_{n,+}(\mathbf{r})] -  n_{\textnormal{F}}[\xi_{n,-}(\mathbf{r})] \Big\}.
\end{equation}

The spin Hall conductivity \eqref{spinHall_conductivity} evidently reduces to the difference between the spin-up and spin-down polarized currents in the limit of vanishing SO coupling and  is completely free of any divergencies for finite SO coupling, even at $T=0$, contrary to the formula for the spin Hall conductivity discussed in Refs. \onlinecite{Shen2004, Shen2005, Bao2005} [we note that Eq. \eqref{Hall_conductivity_full} and consequently Eq. \eqref{spinHall_conductivity} appear to be mathematically very different from the expressions for the conductivities written in these references]. Moreover, the spin Hall current defined in this way is only transverse (only the Hall components of the spin conductivity tensor $\bar{\sigma}^s(\mathbf{r})$ are non-zero) and satisfies by construction a continuity equation  
\begin{equation} 
 \bm{\nabla}_\mathbf{r} \cdot  \mathbf{j}^s_{\textnormal{neq}}(\mathbf{r})=0.
\end{equation}  
As a consequence, we do not have the problems encountered in other definitions of the spin current (see discussion in Ref. \onlinecite{Niu2006} for instance) where the nonconservation of the spin density had to be cured by the introduction of a source term in the continuity equation (which, by the way, has to be present in order to satisfy spin conservation laws).  In addition, the continuity equation allows us to  relate the local (microscopic) spin Hall conductivity to the macroscopic spin Hall conductance, thus hereby providing a physical sense to a macroscopic spin current in the quantum Hall regime (which could be measured by appropriate SO-polarized contacts). Moreover, the present formulation also removes all the ambiguity in the definition of the spin flow which creates persistent equilibrium currents, both in the absence\cite{Rashba2003} and presence of magnetic field, since our angular momentum current density appears only as a response to a nonequilibrium electric field.  

\begin{figure}  
\centering 
\includegraphics[width=0.48\textwidth]{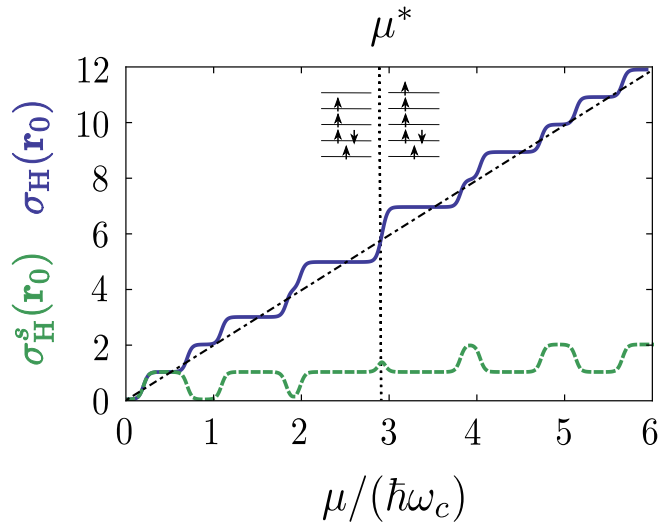}   
\caption{(Color online) Semiclassical Hall and spin Hall conductivity from Eqs. \eqref{Hall_conductivity_full} and \eqref{spinHall_conductivity} (represented by the blue solid and green dashed lines respectively) expressed in units of $e^2/h$ and $e/4 \pi$ as a function of the quantity $\mu/\hbar \omega_c$. Here, SO and Zeeman parameters $S$ and $Z$ are the same as in Fig. \ref{fig1} and the temperature is chosen as $k_B T / (\hbar \omega_c) = 0.03$. The dashed-dotted black line represents the classical result where the Hall conductivity grows linearly with the inverse of the magnetic field. The perpendicular dotted line marks the jump in the Hall conductivity that can be related to an SO imbalance below the Fermi level (as represented by the schema on top).} 
\label{fig4}   
\end{figure} 

In Fig. \ref{fig4}, we compare the local semiclassical charge and spin Hall conductivities plotted as a function of the quantity $\mu/\hbar \omega_c$  for a fixed position $\mathbf{r}_0$ (here the Rashba parameter $S =0.88$ and Zeeman coefficient $Z = -0.37$ correspond to common experimental values in semiconductors). The Hall conductivity exhibits quantized plateaus of different widths related to the fact that the Rashba-Landau energy levels are not equidistant in the energy space. The spin Hall conductivity also exhibits quantized plateaus smeared by temperature whose value oscillates between $0$ and $1$ (in units of $e/4 \pi$) for $\mu \lesssim \mu^\ast$. At the special value $\mu = \mu^\ast$ a jump in the Hall conductivity occurs and, for $\mu \gtrsim \mu^\ast$, the spin Hall conductivity oscillates between $1$ and $2$ due to the imbalance between the two populations of SO resolved states. This imbalance can be related to a crossing in the energy spectrum below the chemical potential where the levels $(n,
+)$ and $(n-1,-)$ have swapped their positions [with $(n,+)$ now energetically lower than $(n-1,-)$]. As such, the energy levels are populated by a different pattern compared to the usual Zeeman spin-split case  (see small inset on top of the figure). The jumps in the Hall conductivity are repeated each time the chemical potential crosses an accidental degeneracy in the (disordered) energy spectrum with an eventual saturation of the SO imbalance for large values of the chemical potential.

\subsection{Macroscopic transport}  
So far, we have not yet discussed macroscopic transport properties in the quantum Hall regime. As emphasized in the Introduction, the conductances turn out to be quite different from the local conductivities at high magnetic fields, because the electronic current density essentially spreads along extended complex structures via a percolation mechanism. Recent scanning tunneling experiments\cite{Morgenstern2008} have confirmed that the local density of states  presents very pronounced spatial inhomogeneities in the quantum Hall regime, with the formation of a percolation cluster of the electronic density in the bulk of the 2DEG at the transitions between Hall plateaus. From a theoretical perspective, a consistent transport theory of the quantum Hall effect requires bringing together the geometric concept of percolation (and fractality) with dissipative processes such as quantum tunneling through the saddle points of the disordered potential landscape, or interactions with phonons.

Such a transport problem constitutes as a whole an important theoretical challenge, which is still topical. In the following we shall just focus on the simpler (albeit still non trivial) high-temperature  regime of the quantum Hall effect, for which a semiclassical transport theory can be formulated\cite{Simon,Floser2011,Floser2012} by neglecting coherence effects. We thus have to deal with a purely classical problem (Landau level quantization is, however, still taken into account), where the objective is to derive the macroscopic current and voltages, starting from the microscopic knowledge of a local Ohm's law ${\bf j}({\bf r})=\bar{\sigma}({\bf r}) {\bf E}({\bf r})$, with the local conductivity tensor
\begin{eqnarray}
\bar{\sigma}({\bf r})=\left(\begin{array}{cc} \sigma_0 & - \sigma_{\textnormal{H}}({\bf r}) \\ \sigma_{\textnormal{H}}({\bf r}) & \sigma_0 \end{array} \right).
\end{eqnarray}
The Hall component $\sigma_{\textnormal{H}}({\bf r})$ is given by formula \eqref{Hall_conductivity_simpl} in the presence of a finite SO interaction, and the uniform diagonal component $\sigma_0$ describes phenomenologically dissipative processes such as electron-phonon scattering. The main difficulty lies in solving the continuity equation ${\bm \nabla}_\mathbf{r} \cdot {\bf j}(\mathbf{r})=0$ by considering a vanishing dissipation $\sigma_0 \to 0$ together with large long-range random spatial fluctuations of the Hall conductivity, i.e., when typically $\langle \delta \sigma_{\textnormal{H}}^2({\bf r}) \rangle \gg \sigma_0^2$ where  $\delta \sigma_\textnormal{H}({\bf r})=\sigma_\textnormal{H}({\bf r}) - \langle \sigma_{\textnormal{H}}({\bf r}) \rangle$ (here we note by $\langle \cdots\rangle$ the spatial average). In other terms, the semiclassical quantum Hall effect problem corresponds to a classical percolation problem within an advection-diffusion regime, which is nonperturbative in nature.  

It has been shown recently\cite{Floser2011,Floser2012} that macroscopic transport coefficients can be captured analytically by an interpolation of the perturbative series expansion [in powers of $\langle \delta \sigma^2_\textnormal{H}({\bf r}) \rangle /\sigma_0^2$] within a general diagrammatic formalism, provided that the Hall conductivity follows a random Gaussian distribution. Given that the spatial fluctuations of $\sigma_\textnormal{H}({\bf r})$ are unaffected by the presence of Rashba SO interaction at the level of semiclassical Eq. \eqref{Hall_conductivity_simpl}, the results of Refs. \onlinecite{Floser2011,Floser2012} established without SO interaction still hold in the case under study in the present paper. Thus, we get the remarkable result that the Hall conductance $G_{\textnormal{H}}$ is independent of the Hall conductivity fluctuations, while the longitudinal conductance $G_L$ encodes altogether dissipation and percolation features. At temperatures $k_B T \gg \sqrt{\langle |V(\mathbf{r})|^2 \rangle}$ such that the fluctuations $\delta \sigma_\textnormal{H}({\bf r})$ are linearly proportional to the disorder potential fluctuations $\sqrt{\langle |V(\mathbf{r})|^2 \rangle}$ [this ensures that the fluctuations  $\delta \sigma_\textnormal{H}({\bf r})$ are Gaussian distributed], we obtain the Hall conductance in the presence of SO interaction
\begin{eqnarray}
G_\textnormal{H}=\frac{e^2}{h} \sum_{n=0}^{+\infty} \sum_{\lambda} n_{\textnormal{F}}\left[E_{n,\lambda} \right],
\end{eqnarray}
where $E_{n,\lambda}$ was originally defined in Eq. \eqref{spectrum}. In addition, we find that the longitudinal conductance is characterized by the same classical percolation exponent $\kappa \approx 0.77$ (the exact value was originally conjectured\cite{Simon} to be $\kappa=10/13$) as in the absence of SO interaction:
\begin{eqnarray}
G_L=C \sigma_0^{1-\kappa} \left|\frac{e^2}{h}\sqrt{\langle |V(\mathbf{r})|^2 \rangle} \sum_{n=0}^{+\infty} \sum_{\lambda} n'_{\textnormal{F}}\left[E_{n,\lambda} \right] \right|^{\kappa}.
\end{eqnarray}
Here $C$ is a nonuniversal constant and $n'_\textnormal{F}(\omega)$ is the derivative of the Fermi-Dirac distribution function.

As a final note, we would like to mention that our finding for the (preserved) classical percolation transport exponent in the presence of SO interaction is fully consistent with previous studies focused on the critical behavior in the quantum Hall regime\cite{Lee1994,Hanna1995,Meir2002}. A change of the universality class is principally expected\cite{Meir2002} for the quantum percolation problem (at zero temperature) when considering SO scattering with short-range correlations which severely spoil coherence of the tunneling events, thus altering the critical percolation exponent. In contrast, smooth forms of disorder\cite{Lee1994,Hanna1995} are not expected to lead to drastic modifications of the percolation transition in the presence of SO interaction.
 
\section{Conclusion}\label{conclusion_sec}

In summary, we have presented in this paper a derivation of the local particle and current densities for a 2DEG in the quantum Hall regime by taking into account the combined presence of a smooth disorder, Rashba SO interaction and Zeeman coupling. To that purpose, we have used a systematic gradient expansion controlled by powers of the magnetic length that allows us to obtain quantum functionals in the local equilibrium or hydrodynamic regime. In a first stage, we have presented the calculation of the leading-order quantum and semiclassical expressions for the electron density and spin polarization. These analytical expressions, accurate for a wide range of temperatures, can be used as a starting point to numerically investigate the effect of electron-electron repulsion at the mean-field level. In addition, both observables reveal clear signatures of the presence of Rashba SO interaction which could be experimentally studied using local probes.

We have also derived within this formalism a general functional expression for the local current density that contains both Landau level mixing processes and quantum effects due to 1D wavefunction delocalization. This local current density exhibits two types of dominant contributions, so-called density-gradient and drift currents, whose spatial distribution reflects the peculiar form of the quantized energy spectrum. As an important result, we have shown, at the semiclassical level, that the microscopic conductivity tensor is quasilocal and the Hall conductivity is quantized in precise integral multiples of the conductance quantum. The value of the Hall plateaus is thus robust to Rashba SO interaction whenever it is finite, its effect being only to change their width. Remarkably, this result also applies to the macroscopic Hall conductance whose quantization remains unaffected, with the longitudinal conductance universal scaling exponent equal to that of the 2DEG without SO coupling.
In the formal (singular) limit of infinite Rashba coupling, which physically describes a decoupled topological insulator surface (i.e., a single Dirac cone) in the presence of smooth disorder, we also prove that the general nontrivial expression for the local Hall component is half-integer quantized due to the nature of the level $n=0$.

Finally, we have shown that the conserved SO-resolved electron current allows us to define a physically meaningful spin Hall conductivity as the difference between the two projections of the SO-resolved components. This spin Hall conductivity does not present any resonances but small jumps related to the spin imbalance below the chemical potential. Future interesting directions of research would be to investigate interaction effects on the basis of our quantum functionals for the local charge and current densities, and to extend the theory for macroscopic transport coefficients towards lower temperatures by keeping track of wave function spreading effects.

\textit{Note added}. Recently, we became aware of some very recent theoretical work (Ref. \onlinecite{Carbotte2014}) studying the quantization of the Hall conductivity in a model of a topological insulator surface in the presence of an additional Schr\"{o}dinger quadratic momentum dispersion. Although we consider exactly the same Hamiltonian as in Ref. \onlinecite{Carbotte2014} and agree on the quantized values of the Hall conductivity in the two limiting cases (pure quadratic dispersion or pure linear momentum dispersion), our conclusions  in the presence of both contributions are strikingly different: Ref. \onlinecite{Carbotte2014} gets deviations to the half-integral or integral quantizations depending on the finite ratio of the Schr\"{o}dinger to the Dirac magnetic energy scales (i.e., on the parameter $S$ within our notation), while we find in our work that the Hall conductivity remains exactly quantized in integral units of $e^2/h$. In other terms, we get the result that the quantization of the Hall conductivity is purely dictated by the topology of the Fermi surface. These different results seem to be rooted in the different formulations of transport: Ref. \onlinecite{Carbotte2014} makes use of a {\em perturbatively based} transport theory (whose validity is questioned  at high magnetic fields  for a disorder with any finite correlation length), while we have derived the Hall conductivity from a {\em nonperturbative} semiclassical transport theory in a smooth electrostatic potential.

\begin{acknowledgments}
Interesting discussions with E. Ya. Sherman and E. J. K\"onig are gratefully acknowledged. D. H.-P. was supported by the RTRA Nanosciences Foundation in Grenoble. This work has also benefited from financial support by ANR ``METROGRAPH'' under Grant No. ANR-2011-NANO-004-06.
\end{acknowledgments}

\appendix

\section{Contribution from $g^{(1)}(\mathbf{R},\omega)$ to the electron current density}\label{appendix_current}
We present in this appendix a more detailed calculation of the quantum expressions for the drift electron current density. We start with the spin-resolved U($1$) contribution which can be obtained by reporting Eq. \eqref{Green_SO_first} into Eqs. \eqref{Glesserdefinition} and \eqref{electron_density_1_general_simplified} along with a summation over the whole frequency spectrum:
\begin{align}\label{electron_density_1_integratedfrequencies}
&\mathbf{j}_{1\sigma}^{\textnormal{dr}}(\mathbf{r}) = \dfrac{e\hbar}{2 m^\ast} \int \dfrac{d^2\mathbf{R}}{2 \pi l_B^2} \sum_{n=0}^{+\infty} \sum_{\lambda_1,\lambda_2} \sum_{\sigma'=\pm} \dfrac{l_B}{\sqrt{2}} \sqrt{n_{\sigma'}+1} \notag \\ &\times 
f_\sigma(\theta_{n}^{\lambda_1})f_{\sigma'}(\theta_{n}^{\lambda_1}) f_\sigma(\theta_{n+1}^{\lambda_2})f_{\sigma'}(\theta_{n+1}^{\lambda_2}) F_{n;\lambda_1;\lambda_2}(\mathbf{R}) \notag \\ &\times   \Big[\mathbf{J}_{n_{\sigma};n_{\sigma}+1}(\mathbf{r},\mathbf{R}) \partial_{+}  + \mathbf{J}_{n_\sigma+1;n_\sigma}(\mathbf{r},\mathbf{R}) \partial_{-}\Big]V(\mathbf{R}).
\end{align}

Here the function $F_{n;\lambda_1;\lambda_2}(\mathbf{R})$ defined in Eq. \eqref{Gamma_function} contains the Fermi-Dirac distribution functions for adjacent Rashba-Landau energy levels [as a consequence of the level mixing induced by $\partial_{\pm}V(\mathbf{R})$] and $\mathbf{J}_{n_1;n_2}(\mathbf{r},\mathbf{R})$ is the current kernel \eqref{current_kernel}. We also remind the reader that the partial derivatives are defined as $\partial_{\pm}=\partial_X \pm i \partial_Y$.

Dominant contributions come either from the diagonal elements of the electronic kernel $K_n(\mathbf{r}-\mathbf{R})$ or from terms differing in two Landau levels $K_{n; n\pm 2}(\mathbf{r},\mathbf{r},\mathbf{R})$. The latter type of electronic kernels satisfy the identities\cite{Champel2008}
\begin{align}\label{appendix_relation_kernels_1}
 \sqrt{n+2}\sqrt{n+1} & K_{n;n+2}(\mathbf{r},\mathbf{r},\mathbf{R}) = \notag \\ &  \dfrac{l_B^2}{2}\sum_{q=0}^{n}(n+1-q)\partial_{-}^2K_q(\mathbf{r}-\mathbf{R}),
\end{align}
\begin{align}
 \dfrac{l_B^2}{2}\Delta_\mathbf{r}\sum_{q=0}^{n}(n+1-q)K_q(\mathbf{r}-\mathbf{R})&=(n+1)K_{n+1}(\mathbf{r}-\mathbf{R}) \notag \\ &-\sum_{q=0}^{n}K_q(\mathbf{r}-\mathbf{R}),
\end{align}
so that the bracketed term in Eq. \eqref{electron_density_1_integratedfrequencies} can be rewritten in the form
\begin{multline}\label{appendix_relation_kernels_2}
 \dfrac{l_B^2}{2}\begin{pmatrix}
                  \textnormal{Im}\\
                  \textnormal{Re}
                 \end{pmatrix}
\partial_{+}V(\mathbf{R}) \sum_{q=0}^{n_\sigma}(n_\sigma+1-q)\partial^2_{-}K_{q}(\mathbf{r}-\mathbf{R})  =\\\hat{\mathbf{z}} \times \bm{\nabla}_\mathbf{R} V(\mathbf{R}) \left[\sum_{q=0}^{n_\sigma} K_{q}(\mathbf{r}-\mathbf{R}) -(n_\sigma+1)K_{n_\sigma+1}(\mathbf{r}-\mathbf{R})  \right]  \\
 +l_B^2 \hat{{\bf z}} \times  \left[\bm{\nabla}_\mathbf{R} V(\mathbf{R}) \cdot  \bm{\nabla}_\mathbf{r} \right] \bm{\nabla}_\mathbf{r} \sum_{q=0}^{n_{\sigma}} (n_{\sigma}+1-q) K_{q}(\mathbf{r}-\mathbf{R}).
\end{multline}
 Using also that 
\begin{eqnarray}
 \bm{\nabla}_\mathbf{r} \left[  \bm{\nabla}_\mathbf{R}V \cdot  \bm{\nabla}_\mathbf{r} K_q(\mathbf{r}-\mathbf{R}) \right]=\left[\bm{\nabla}_\mathbf{R}V \cdot  \bm{\nabla}_\mathbf{r} \right]  \bm{\nabla}_\mathbf{r} K_q(\mathbf{r}-\mathbf{R}) \nonumber,
\end{eqnarray}
the final result for the U$(1)$ contribution can then be written as in Eq. \eqref{current_U1}.

For the SU$(2)$ part we proceed in the same way, inserting Eq. \eqref{Green_SO_first} into Eqs. \eqref{Glesserdefinition} and \eqref{electron_density_2_general} and integrating over the frequencies to obtain
\begin{align}\label{electron_density_2_integratedfrequencies}
\mathbf{j}_{2\sigma}^{\textnormal{dr}}(\mathbf{r}) &= \dfrac{i\alpha e l_B}{\sqrt{2}} \int \dfrac{d^2\mathbf{R}}{2\pi l_B^2}\sum_{n=0}^{+\infty} \sum_{\lambda_1,\lambda_2} \sum_{\sigma'=\pm} \sqrt{n_{\sigma'}+1} \notag \\ &\times f_{\sigma}(\theta_{n}^{\lambda_1}) f_{\sigma'}(\theta_{n}^{\lambda_1})f_{-\sigma}(\theta_{n+1}^{\lambda_2})f_{\sigma'}(\theta_{n+1}^{\lambda_2})  \notag \\ &\times F_{n;\lambda_1;\lambda_2}(\mathbf{R}) \Big[K_{n_\sigma;n_{-\sigma}+1}(\mathbf{r},\mathbf{r},\mathbf{R})\partial_{+}\, \notag \\ &(\mp)\, K_{n_\sigma;n_{-\sigma}+1}(\mathbf{r},\mathbf{r},\mathbf{R})\partial_{-} \Big] V(\mathbf{R})
 \begin{pmatrix}
\sigma \\
i
\end{pmatrix},
\end{align}
where $(\mp)$ means taking $-$ for the $\hat{\mathbf{x}}$ component and $+$ for the $\hat{\mathbf{y}}$ component. As happened for the SU$(2)$ density-gradient current density, we have to perform the sum over the two spin-projected currents to combine the pairs of electronic kernels appearing in Eq. \eqref{electron_density_2_integratedfrequencies} between square brackets and obtain a real current. As a result, one also gets pairs of kernels that are either diagonal in the index $n$ or differ by two levels $n,n\pm2$. The terms coupling non-adjacent Rashba-Landau energy levels [see Eqs. \eqref{appendix_relation_kernels_1}-\eqref{appendix_relation_kernels_2} above] do not contribute to the current in the strict semiclassical limit, contrary to the U$(1)$ part. Finally, collecting all the contributions, we get the result quoted in Eq. \eqref{current_U2}.


\begin{thebibliography}{99}
\bibitem{footnote1}
We shall assume in this article that $1-Z >0$ in agreement with typical experimental values of the $g$ factor in semiconductors.
\bibitem{footnote2}
We note that for the Landau level $n= 0$, Eq. \eqref{angles} is perfectly well-defined since $\textnormal{lim}_{n \rightarrow 0}\, \theta_n^{-} = 0$.
\end{thebibliography}
\end{document}